\documentclass[preprint,12pt]{aastex}
\usepackage{emulateapj5}
\usepackage{amsmath}
\usepackage{natbib}

\begin{document}


\def\mbh{{$\mathcal M_{\rm BH}$}}
\def\msol{{$\mathcal M_{\odot}$}}
\def\mr{{$M_R$}}
\def\lr{{$L_R$}}
\def\lbulge{{$L_{\rm bulge}$}}
\def\msig{{$\mathcal M_{\rm BH}$-$\sigma_*$}}
\def\mbulge{{$\mathcal M_{\rm bulge}$}}

\shortauthors{PENG ET AL.}
\shorttitle{BLACK HOLE AND BULGE CO-EVOLUTION}

\title{PROBING THE COEVOLUTION OF SUPERMASSIVE BLACK HOLES AND QUASAR
HOST GALAXIES\altaffilmark{1}}

\author {Chien Y. Peng\altaffilmark{2,3,4}, Chris D.  Impey\altaffilmark{4},
Luis C. Ho\altaffilmark{5}, Elizabeth J. Barton\altaffilmark{4,6,7}, \&
Hans-Walter Rix\altaffilmark{8}}

\altaffiltext{1} {Based on observations with the NASA/ESA {\it Hubble Space
Telescope}, obtained at the Space Telescope Science Institute, which is
operated by AURA, Inc., under NASA contract NAS5-26555.}

\altaffiltext{2}{Current address: Space Telescope Science Institute,
3700 San Martin Drive, Baltimore, MD 21218; cyp@stsci.edu.}

\altaffiltext{3}{STScI Fellow}

\altaffiltext{4}{Steward Observatory, University of Arizona, 933 N. Cherry
Av., Tucson, AZ 85721;  cimpey@as.arizona.edu.}

\altaffiltext{5}{The Observatories of the Carnegie Institution of Washington,
813 Santa Barbara St., Pasadena, CA 91101; lho@ociw.edu.}

\altaffiltext{6}{Hubble Fellow}

\altaffiltext{7} {Current address: University of California at Irvine,
Department Physics and Astronomy, 4154 Frederick Reines Hall, Irvine, CA
92697; ebarton@uci.edu.}

\altaffiltext{8}{Max-Planck-Institut f\"{u}r Astronomie, K\"onigstuhl 17,
Heidelberg, D-69117, Germany; rix@mpia.de}

\begin {abstract}

At low redshift, there are fundamental correlations between the mass of
supermassive black holes (\mbh) and the mass (\mbulge) and luminosity of the
host galaxy bulge.  We investigate the same relation at $z\gtrsim1$.  Using
virial mass estimates for 11 quasars at $z\gtrsim2$ to measure their black
hole mass, we find that black holes at high-$z$ fall nearly on the same \mbh\
versus $R$-band magnitude (\mr) relation (to $\sim 0.3$ mag) as low-redshift
active and inactive galaxies, {\it without} making any correction for
luminosity evolution.  Using a set of conservative assumptions about the host
galaxy stellar population, we show that at $z\gtrsim 2$ (10 Gyr ago) the
ratio of \mbh/\mbulge\ was 3--6 times larger than today.  Barring unknown
systematic errors on the measurement of \mbh, we also rule out scenarios in
which moderately luminous quasar hosts at $z\gtrsim2$ were fully formed bulges
that passively faded to the present epoch.  On the other hand, 5 quasar hosts
at $z\approx1$ are consistent with current day \mbh-\mr\ relationship after
taking into account evolution, appropriate for that of E/S0 galaxies.
Therefore, $z\approx1$ host galaxies appear to fit the hypothesis they are
fully formed early-type galaxies.  We also find that most quasar hosts with
absolute magnitudes brighter than \mr\ = $-23$ cannot fade below $L_*$
galaxies today, {\it regardless} of their stellar population makeup, because
their black hole masses are too high and they must arrive at the local
\mbh-\mr\ relationship by $z=0$.


\end {abstract}


\section {INTRODUCTION}

There are currently $\sim40$ galaxies in which a supermassive black hole (BH)
has been directly measured using stellar and gas dynamics (e.g., reviews by
Barth 2004 and Kormendy 2004).  With a large number of secure detections,
several correlations have been discovered between the mass of the BHs and the
global properties of the galaxy bulge, such as its stellar velocity dispersion
($\sigma_*$), mass (\mbulge), and luminosity (e.g., Kormendy \& Richstone
1995; Magorrian et al.  1998; Ho 1999; Gebhardt et al.  2000a; Ferrarese \&
Merritt 2000; Kormendy \& Gebhardt 2001, H\"aring \& Rix 2004).  What makes
the correlations even more striking is the relatively small amount of
intrinsic scatter; the relation between \mbh\ and $\sigma_*$ (Gebhardt et al.
2000a, and Ferrarese \& Merritt 2000) has only a scatter of $\lesssim2.2$ (0.3
dex; Tremaine et al.~2002) in \mbh.  In the original discovery, the
relationship between \mbh\ and the bulge luminosity appeared weaker, with a
scatter of 0.5 dex (Magorrian et al.  1998, Kormendy \& Gebhardt 2001).  When
the bulge luminosity (\lbulge) is translated into mass, one finds that the
ratio of the BH mass to the bulge mass is \mbh/\mbulge\ $\approx$ 0.0012
(Kormendy\& Gebhardt 2001, Merritt \& Ferrarese 2001, and McLure \& Dunlop
2002). These remarkable findings leave little room to doubt that the formation
and evolution of BHs and galaxy bulges are closely tied.  Understanding why
provides fundamental insights toward a coherent understanding of galaxy
formation and evolution in general.


To study the coevolution of black holes with galaxy bulges, there is a great
deal of interest in investigating the correlations at higher redshifts.  In
this study, we explore the \mbh\ versus the $R$-band luminosity (\mr) relation
of the galaxy bulge out to $z\approx 2$.  The \mbh-\mr\ correlation has been
studied by McLure \& Dunlop (2002), and Bettoni et al.  (2003) for $z\le1$
galaxies, and is an extension of the studies pioneered by Kormendy \&
Richstone (1995) and Magorrian et al.~(1998).  Since Kormendy \& Richstone's
(1995) original finding, which was based on $B$-band images of nearby
galaxies, there have been many efforts to sharpen the correlation using larger
samples and/or redder passbands (Magorrian et al.  1998; Ho 1999; Kormendy \&
Gebhardt 2001; Laor 2001; Merritt \& Ferrarese 2001; McLure \& Dunlop 2002;
Bettoni et al.  2003; Marconi \& Hunt 2003, Ivanov \& Alonso-Herrero 2003).
The RMS scatter of the correlation, when both early- and late-type galaxies
with \mbh\ measurements are taken into account, is fairly large ($\sim 0.5$
dex; Kormendy \& Gebhardt 2001; Marconi \& Hunt 2003).  When attention is
focused on just the early-type galaxies, several studies find the scatter in
the \mbh-bulge luminosity relation drops to 0.3-0.45 dex (Erwin, Graham, \&
Caon 2002, McLure \& Dunlop 2002; Bettoni et al. 2003; Marconi \& Hunt 2003).
While it is not clear what causes the larger scatter in late-type galaxies,
one of the complications is that it is not always straightforward to obtain
bulge luminosities, which is hard to uniquely separate from the disk in
late-type galaxies when the bulge does not have a classical de Vaucouleurs
profile (e.g.  Carollo 1999).

While measuring the bulge luminosity of {\it normal} (i.e.  inactive) galaxies
is feasible out to high $z$, measuring the mass of their central black holes
accurately is considerably more difficult.  Currently, the most robust method
for measuring \mbh\ in normal galaxies is through modeling the stellar and gas
dynamics (e.g., Barth 2004 and Kormendy 2004).  Doing so requires an exquisite
spatial resolution on the order of the BH sphere of influence, $r_{\rm sph} =
G {\mathcal M}_{\rm BH}/\sigma_*^2$, the radius at which the Keplerian orbital
velocity of the stars due to the influence of the central BH is comparable to
the stellar velocity dispersion of the bulge.  This is feasible only for the
nearest, and often relatively luminous, galaxies.  The \msig\ relation in
nearby galaxies offers an easier way to infer \mbh.  However, despite the
potential for extending this technique widely to infer \mbh\ to study bulge
and BH evolution, in practice the technique is time consuming and has
practical challenges:  $\sigma_*$ is difficult to measure in a variety of
circumstances, including low-mass and low-surface brightness galaxies, distant
galaxies, and galaxies in which the active nucleus substantially dilutes the
starlight (Greene \& Ho 2005a).  More fundamentally, to study the {\it
co}-evolution of BHs with galaxy bulges, it is desirable to measure \mbh\
using a technique that is more directly associated with the influence of a
black hole, especially at high $z$, rather than one which is tied to the
bulge.

On the other hand, active galactic nuclei (AGNs), whose energetic output is a
direct manifestation of supermassive black holes, are natural and promising
candidates for measuring \mbh, using a technique known as the ``virial''
technique (e.g. see Ho 1999; Wandel, Peterson, \& Malkan 1999).  This
technique is based on measuring AGN broad-line region (BLR) sizes and
linewidths that have been subjected to reverberation mapping (Blandford \&
McKee 1982; Peterson 1993).  These simple \mbh\ estimates appear to obey the
\msig\ relation of inactive galaxies at low redshift ($z\lesssim 0.05$)
(Gebhardt et al.  2000b; Ferrarese et al.  2001; Nelson et al. 2004, Onken et
al. 2004; Barth, Greene, \& Ho 2005; Green \& Ho 2005b), which provides an
important cross-check on the validity of the AGN virial masses.  For a sample
of 34 low-$z$ Seyfert 1 nuclei and quasars with reverberation mapping data for
the H$\beta$ line, Kaspi et al.  (2000) showed that the BLR size correlates
strongly with the AGN continuum luminosity.  This provides a convenient
shortcut to estimate BH masses essentially for any (unbeamed) broad-line AGN
with a rest-frame optical spectrum.  It offers a potential to extend the
measurement of \mbh\ out to a large sample of objects, spanning a wide range
in redshift.  Since then, this virial mass measurement technique has been
bootstrapped to higher redshifts using the ultraviolet lines C~{\sc iv}
$\lambda$1549~\AA\ (Vestergaard 2002) and Mg~{\sc ii} $\lambda$2798~\AA\
(McLure \& Jarvis 2002).  The calibration sample, however, is small, currently
based on $\lesssim 60$ objects.

Using the virial mass technique, the measurement of supermassive BHs has now
been extended out to large redshifts in a large number of quasars.  Their
numerical abundance at high redshift ($z>1$) and their high luminosities make
them ideal primary tracers for BH evolution, thus their host galaxies prime
candidates for studying galaxy evolution.  The trade-off for having an easy
proxy for measuring \mbh\ in quasars is that detecting quasar host galaxies is
a much harder task compared with studying inactive galaxies at high $z$ (i.e.,
$z\gtrsim1$).  The difficulty in extracting robust parameters for the host
galaxy in the glare of a luminous quasar is well known.  It is worth pointing
out that even with the exquisite 0\farcs05 resolution images of the {\it
Hubble Space Telescope (HST)} in the optical, the task remains nontrivial
because of the small host galaxy size and its low surface brightness, which
decreases as $(1+z)^4$.  Despite technical challenges, careful observations
are detecting more and more quasar hosts at high $z$, even from the ground,
using deep imaging and adaptive optics (Fynbo, Burud, \& M{\o}ller 2002; Lacy
et al.  2002; Hutchings 2003; S\'anchez \& Gonz\'alez-Serrano 2003, Kuhlbrodt
et al. 2005).

In this study we explore the \mbh\ vs. \mr\ relationship for quasar host
galaxies out to $z\gtrsim1$.  We use the virial technique to estimate \mbh\ in
quasars, combined with $R$-band luminosity of the bulge of quasar host
galaxies, inferred from {\it HST}\ studies of Kukula et al.~(2001), Ridgway et
al.~(2001) and Peng et al.~(2005).  At low $z$ quasars brighter than
$M_V=-23.5$ primarily live in elliptical galaxies (McLeod \& Rieke 1995 and
Dunlop et al.  2003), and there are hints that indicate hosts at $z\gtrsim1$
may be early-types as well.  In particular, Kukula et al.  (2001) and
S\'anchez \& Gonz\'alez-Serrano (2003) find that the luminosities and scale
lengths of galaxies are consistent with a passively evolving scenario, and,
moreover, that they follow the Kormendy relation (the correlation between
half-light surface brightness $\mu_{1/2}$ and effective radius $r_{1/2}$).
Therefore, in this study, we also explore the evolution in the
\mbh-\mr\ relationship by testing the null-hypothesis that the host galaxies
evolve like E/S0 galaxies from a formation redshift of $z_f=5$.

We note that at low redshift, the scatter in the \mbh-\mr\ relationship for
elliptical galaxies (McLure \& Dunlop 2002 and Bettoni et al.~2003) is small
enough that when extended to high redshift it can provide a sensitive and
useful probe of the coevolution of BH masses with bulges.  Black holes are a
useful constraint for understanding galaxy evolution because \mbh\ only
increases, which also makes it a way to age-date a galaxy.  Coupled to
\lbulge, galaxies at any given redshift are restricted in how much they can
fade in the two-dimensional diagram of \mbh\ vs. \lbulge.  They cannot freely
roam the \mbh-\lbulge\ diagram, limited on the one hand by the monotonicity of
\mbh, on the other by the requirement that they evolve toward the tight local
\mbh\ vs. \lbulge\ relation at $z=0$, and additionally by the local galaxy
luminosity function and the mass function of supermassive black holes.  Thus,
at a given bulge mass, luminosity, and redshift, galaxies may follow a range
of possible merger and luminosity evolution paths that are initially wide but
progressively narrow as they evolve towards the tight local relations of \mbh\
versus luminosity.  Generalizing the idea into $n$-dimensional space, by
measuring the \mbh\ vs. luminosity relation in several passbands, this
``phase-diagram'' may indicate the state of galaxies at high redshift relative
to today.  This may ultimately be a powerful, yet simple way to visualize and
understand the complex paths of galaxy formation and evolution.  The technique
requires calibrating \mbh\ vs.  luminosity at many passbands for local
galaxies.

Our discussion below is structured in the following order.  To obtain the
\mbh-\mr\ relation for quasar host galaxies, one of the first steps is to make
{\it k}-correction to the host galaxy photometry (\S~2).  In Section 3, we
will then estimate the BH masses using the virial technique, compare them to a
cruder estimate using the \mbh-luminosity relation for quasars determined by
Peterson et al.\ (2004), and discuss the Eddington efficiencies of the quasars
in the sample, using black hole masses estimated from the virial technique.
Section 4 presents the results, and we conclude with brief remarks (\S~5).  In
this study, we use a standard cosmology with $H_0=70$ km$^{-1}$ s$^{-1}$
Mpc$^{-1}$, $\Omega_{\rm m} = 0.3$, and $\Omega_\Lambda = 0.7$.

\section {DATA and BULGE {\it k}-CORRECTION}

We assemble a set of data on quasar host galaxies published by Kukula et al.
(2001), Ridgway et al. (2001), and Peng et al. (2005) observed using {\it
HST}/NICMOS, summarized in Table 1.  All the magnitudes shown in the Table
have been corrected for Galactic extinction according to Schlegel, Finkbeiner,
\& Davis (1998), but not internal extinction.  The Kukula et al. (2001) sample
consists of 18 quasars at $z\approx1$ and $z\approx2$, equally divided between
radio-loud quasars (RLQ) and radio-quiet quasars (RQQ), while the Ridgway et
al. (2001) sample consists of 5 RQQs at $1.5 \le z \le 2.8$.

The Peng et al.  (2005) study has one additional object, obtained using the
gravitational lensing technique of the CASTLES survey (Kochanek et al. 1999).
In that study, they find that the host galaxy has been stretched out into a
luminous Einstein ring.  After deprojection, where the lensing galaxy mass is
assumed to be a singular isothermal ellipsoid, they recovered the intrinsic
light distribution of the host galaxy (e.g. size, luminosity, axis ratio,
etc.).  The host galaxy Einstein ring in CTQ~414 is only modestly merged with
the lensing galaxy light profile.  Even so, because the shape of the host
galaxy Einstein ring is so dramatically different azimuthally from the lensing
galaxy, it is fairly easy to disentangle the two.  They did so by modeling the
lens deflection model, light profile of the lens, the host, and the quasar
images -- all simultaneously.  In so doing, they find that the uncertainty of
the host luminosity is roughly 0.2 mag.  As we shall see, the result for this
object would only start to be important if it were one magnitude {\it
brighter} than our number.  Even so, individually, no one object will
significantly affect the overall conclusion.  Thus this object alone serves
mostly as a consistency check, as it is the most well-resolved object.

The data from all samples were observed in either NICMOS F110M ($\sim J$ band)
or F160W/F165M ($\sim H$ band) filters so that they roughly correspond to
rest-frame $V$, where the contrast between the host and the quasar is
favorable for detecting the host galaxy.  Altogether, $24$ host galaxies are
used in this study, which is the entire collection of data in the literature
published specifically to study quasar host galaxies detected using NICMOS
above $z\gtrsim1$.  There are other host galaxies found at $z>2$ published
using WFPC2 and ground-based data (e.g., Aretxaga, Terlevich, \& Boyle 1998;
Lehnert et al. 1999; Hutchings et al.  2002; Hutchings 2003).  We exclude
these because the observations were made in the rest-frame UV, which is very
sensitive to star formation and extinction internal to the host galaxy.
Therefore, $k$-correcting to the rest-frame $R$ filter is highly dependent on
assumptions of the spectral energy distribution (SED) of the galaxy.  We also
exclude ground-based data of RLQ hosts published by Falomo, Kotilainen \&
Treves (2001), Lacy et al. (2002), S\'anchez \& Gonz\'alez-Serrano (2003), and
Kuhlbrodt et al. 2005, to keep the sample uniform and to avoid potential
uncertainties with the quasar-host separation relative to data obtained
uniformly with {\it HST}.

To infer the $R$-band absolute magnitude (\mr) of the quasar host galaxy bulge
from the raw values published in Kukula et al.  (2001), Ridgway et al.
(2001), and Peng et al. (2005), we make several corrections.  We note that the
published host galaxy fluxes primarily refer to the bulge because the images
are too shallow to detect the diffuse components of extended disks if they
exist, except for CTQ~414 (Peng et al. 2005).  For that object, we take the
bulge luminosity only.  We point out that since the local black hole versus
bulge relation excludes the disk component, as we shall see, a proper removal
of a disk would further strength our conclusions by making the bulges fainter.
Ridgway et al. (2001) compared their host galaxy detection with simulations,
from which they determined the host galaxy flux as a function of the
morphological type.  Thus, for their sample, to obtain the $H$-band bulge
luminosity shown in Table 1, we apply an aperture correction (their Table 4)
to their 1\farcs01 aperture photometry (their Table 2), and a correction that
is dependent on assumptions about the morphology of the host (their Table 5),
appropriate for radio galaxies, which have the largest corrections in their
study.  Finally, to convert published photometry into a rest-frame, standard
Cousins, $R$-band flux, we follow Hogg et al.  (2004) in computing detailed
$k$-corrections.  We transform the {\it HST} Vega-based magnitudes to $R$-band
by using galaxy templates observed by Coleman, Wu, \& Weedman (1980),
multiplied by the appropriate filter transmission curves.  The integrated
fluxes are normalized to a spectrum of $\alpha$-Lyr, ``observed'' in the
appropriate bandpass.

Computing $k$-corrections to the host galaxy's rest-frame $R$-band relies on
having an SED.  We have no color information for most of the host galaxies so
we must make an assumption.  The filters used by Kukula et al.  (2001; $J$ and
$H$) and Ridgway et al.  (2001; $H$) conveniently capture the hosts in their
rest-frame $V$ (except for two objects at $z\approx 2.7$ in Ridgway et al.
2001).  Therefore, $k$-correction to rest-frame $R$ is not very sensitive to
assumptions of the SED for both the quasar or the host galaxies, especially as
the $V$-filter lies redward of the rest-frame 4000~\AA\ break in galaxies.
Specifically, the difference in the inferred \mr\ between an early and
late-type SED is only $\sim 0.1$ mag at $z=1$ and $\sim0.3$ mag at $z=2$.  For
the $k$-correction we use an SED of {\it current-day} E/S0s for the host
galaxies because it is the most conservative assumption with regard to the
conclusions we are testing (i.e. weakening it), despite it being less sensible
than other potential models.  All other more realistic SEDs at $z\sim2$ are
likely to be bluer, thus the inferred host luminosities would be
systematically fainter than the \mr\ values shown in Table 2.


\section {BLACK HOLE MASS ESTIMATE and EDDINGTON RATIO}

\subsection {Virial \mbh\ Measurement}

For high redshifts, the virial technique relies on measuring the quasar
continuum luminosity and the widths (FWHM) of the C~{\sc iv}
($\lambda$1549~\AA) or Mg~{\sc ii} ($\lambda$2798~\AA) emission line.  At
$z\sim1$ the Mg~{\sc ii} emission line is the most convenient feature to use
because it falls into optical bandpasses where the atmosphere is transparent
and where CCDs are sensitive.  Likewise, at $z\gtrsim2$ C~{\sc iv} is the line
of choice.  Compared to the H$\beta$ line, C~{\sc iv} is produced much closer
to the supermassive black hole (by factor of $\sim 0.5$).  As such there are
concerns that the kinematics of the BLR, not well understood, may be dominated
by non-Keplerian motions such as winds and outflows.  Concerns about the UV
emitting clouds being caught up in a disc also meant that the width of the
emission lines may depend strongly on the viewing angle.  Despite
uncertainties in the BLR physical structures, the same virial product that is
used to estimate black hole masses in AGNs using H$\beta$ is found to have
analogs in the UV with C~{\sc iv} or Mg~{\sc ii} lines, over the luminosity
range present in our AGN sample (Vestergaard et al.  2002, McLure \& Jarvis
2002, Kaspi et al.  2005).  The assumption we will make is that the same
technique applies to quasars at high redshifts.

The high redshift calibration for C~{\sc iv} was performed by Vestergaard
(2002)\footnote{She used a cosmology of $H_0=74$ km s$^{-1}$ Mpc$^{-1}$,
$\Omega_\Lambda=0$, and $\Omega_{\rm m}=1$.}.  In viral estimates of \mbh\
there is a geometric normalization factor, $f$, that depends on the structure
and kinematics of the BLR, and which normalizes the AGN \msig\ to that of the
quiescent galaxies.  This factor was determined empirically by Peterson et al.
(2004) and Onken et al.  (2004), to be $\left<f\right>=5.5$ for the H$\beta$
BLR (under an assumption that $\sigma_{line} =$~FWHM$/2$).  In the absence of
a large sample to determine $f$, the effective normalization used in Kaspi et
al. (2000) corresponded to $\left<f\right> = 3$, so that the new \mbh\
estimates are a factor of $5.5/3=1.8$ higher than previously estimated in
Kaspi et al.  (2000).  With these caveats, M.~Vestergaard (2005, private
communication) provided the following relation to estimate \mbh\ from C~{\sc
iv} ($\lambda$1549~\AA), appropriate for our adopted cosmology:

\begin {equation}
{\mathcal M}_{\rm BH} = 4.5\times10^6\left[\frac{{\rm FWHM(\mbox{C~\sc iv})}}{1000\ {\rm km\ s}^{-1}}\right]^2\left[\frac{\lambda L_\lambda(1350\mbox{~\AA})}{10^{44}\ {\rm erg\ s}^{-1}}\right]^{0.53} \, {\mathcal M}_{\odot},
\end {equation}

\noindent where $L_\lambda(1350\mbox{~\AA})$ is the spectral density of the
quasar continuum at rest-frame 1350~\AA.  The value of the exponent
($\gamma=0.53$) and the coefficient (4.5$\times10^6$) depend on the sample
used in the fit and the fitting technique (BCES [Akritas \& Bershady 1996]
vs.\ FITEXY [Press et al. 1992]).  There may be small systematic differences,
but they are otherwise formally equivalent to within the errors (Kaspi et al.\
2005).

For $z\approx1$ quasars Mg~{\sc ii} ($\lambda$2798~\AA) is a good surrogate
for estimating \mbh\ that conveniently falls in the visible region (McLure \&
Jarvis 2002).  Again, we renormalize their empirical relation by a factor of
$f=1.8$ to obtain:

\begin {equation}
{\mathcal M}_{\rm BH} = 6.1\times10^6\left[\frac{{\rm FWHM(\mbox{Mg~\sc ii})}}{1000\ {\rm km\ s}^{-1}}\right]^2\left[\frac{\lambda L_\lambda(3000\mbox{~\AA})}{10^{44}\ {\rm erg\ s}^{-1}}\right]^{0.47} \, {\mathcal M}_{\odot}.
\end {equation}

To obtain the C~{\sc iv} and Mg~{\sc ii} emission line FWHM in quasars with
host detections, we searched the literature for published spectra.  In total,
we compile 16 (out of a total of 24) objects that have published spectra or
estimated FWHM values -- 11 at $z\gtrsim 2$ and 5 at $\sim 1$.  The Ridgway et
al.  (2001) sample comes from a survey published by Zitelli et al.  (1992),
while the Kukula et al.  (2001) sample is a composite of different surveys
shown in Table 1.  For CTQ~414, we obtained an uncalibrated spectrum at the
Magellan telescope.  We note that the quasar spectra published in Boyle et al.
(1990) used by Kukula et al.  (2001) sample have very uncertain flux
calibrations, and likewise for CTQ~414.  However, what matters more to our
analysis is the FWHM values, given that the absolute fluxes are determined by
photometry (see below).  Moreover, because the continuum luminosity enters
into Eqs.~1 and 2 at most as the power of 0.53 (0.47 for Mg~{\sc ii}), even a
flux uncertainty as large as a factor of 2 translates into an uncertainty in
\mbh\ of at most 0.2 dex, which will not affect our conclusions.  To measure
the widths of the emission lines that do not have published values, two of the
authors independently measured the FWHM manually from compiled spectra, in a
double-blind test.  The two independent measurements of the FWHM agree well.
One other potential uncertainty in measuring the FWHM is that the broad
emission lines may be contaminated by Fe~{\sc ii} emission (Vestergaard \&
Wilkes 2001).  Fe~{\sc ii} contamination is more important for H$\beta$ and
Mg~{\sc ii} lines (Vestergaard \& Wilkes 2001), and much less for C~{\sc iv}.
For the Mg~{\sc ii} line we do not have a way to estimate the Fe~{\sc ii}
contribution, thus we do not correct for this effect.  This is appropriate for
the calibration performed by McLure \& Jarvis (2002), which did not remove the
Fe~{\sc ii} contamination as well.  All of our $z\gtrsim 2$ objects use C~{\sc
iv} measurements and only our $z\approx 1$ objects rely on Mg~{\sc ii} lines.

To estimate the monochromatic luminosity of the quasars at
$L_\lambda(1350\mbox{~\AA})$ and $L_\lambda(3000\mbox{~\AA})$, we first
$k$-corrected to restframe $U$-band, centered on 3745~\AA, using a powerlaw
SED that fits to apparent magnitudes in $H/J$, and $B/V$ filters.  The same
SED is then used to extrapolate to 1350~\AA\ or 3000~\AA\ as required.  To
convert from $U$-band magnitude to flux density we use the spectral irradiance
value of $4.34\times10^{-9}$ erg s$^{-1}$ cm$^{-2}$~\AA$^{-1}$ (Colina,
Bohlin, \& Castelli 1996).  There is a small conversion to take the $B$-band
magnitudes from literature, based on photographic plates (see Table 1), into
the standard Vega-magnitude system.  The magnitudes of the SGP objects from
Boyle et al.  (1990) were based on plate-scanned $B_J$ system (``$b$-band''),
zeropointed to the Johnson $B$ magnitude system.  The conversion to the
standard $B$ magnitude has a small offset for typical quasar colors:  $b = B +
0.1\pm 0.05$.  In comparison, the random errors are generally larger, about
0.2-0.3 mag.  Similarly, the magnitude of the MZZ objects from Zitelli et al.
(1992) is based on photographic $B_J$ filter system.  As the systematic
corrections are small and will not affect our conclusions, we do not apply a
correction.  We correct the $B$-band magnitudes for Galactic extinction,
according to Schlegel, Finkbeiner, \& Davis (1998).

This technique for measuring the quasar flux density near 1350~\AA\ and
3000~\AA\ is fairly robust, as the $B/V$ and $J/H$ filters cover a long
baseline for estimating a powerlaw SED.  Since the observed $B$ band also
falls in the restframe UV, there is only a small amount of SED-dependent
extrapolation.  We compare \mbh\ estimates using our SED extrapolation with
another crude method of directly converting observed $B$-band and $V$-band
fluxes into \mbh.  Both methods give nearly the same results.  Contamination
by emission lines in the broad-band filter is often negligible even in quasars
because the bandwidths of the $V$ (FWHM = 890~\AA) and $B$ (FWHM=1000~\AA)
filters are significantly larger than the widths of the lines (FWHM
$\lesssim\, 100$~\AA).  The impact of such errors on \mbh\ is negligible
($\lesssim 5\%$) because the monochromatic luminosity enters into Eqs.~1 and 2
as the powers of 0.53 and 0.47, respectively.  All the transformed values are
shown in Table 2.  For CTQ~414 and 4C~45.51 we take the average of \mbh\
estimates from the C~{\sc iv} and Mg~{\sc{ii}} techniques.

Lastly, we point out that with only two objects, 4C~45.51 and CTQ~414, that
have both C~{\sc iv} and Mg~{\sc ii} simultaneously observed it is not
possible to check on systematic differences in the \mbh\ estimates:  in
4C~45.51, the inferred black hole mass is comparable between using C~{\sc iv}
($0.65\times 10^9$\msol) and Mg~{\sc ii} ($0.58\times10^9$\msol).  In the
other, CTQ~414, C~{\sc iv} yields a factor of 2 larger BH mass than Mg~{\sc
ii}.  However, in a separate study using gravitationally lensed quasars at
$z\gtrsim 1$ (Peng et al. in prep), including 4 other objects that have
simultaneous C~{\sc iv} and Mg~{\sc ii} observations, the two methods appear
not to have a systematic offset.

\subsection {\mbh-Luminosity (L$_{QSO}$) Relation of Quasars}

Using virial estimates of \mbh, it has been pointed out that, to first order,
BH masses correlate with the unresolved AGN optical/UV luminosity (e.g.  see
Kaspi et al.  2000; Oshlack, Webster, \& Whiting 2002; Peterson et al.  2004).
This correlation is an empirical statement that while the distribution of
Eddington ratio $\epsilon$ in AGNs is large, there might be a preferential
efficiency over an average lifetime once they are ``turned on.''  This typical
$\epsilon$ might also depend on the luminosity range of AGN sample under
consideration.  For nearby AGNs the mass-luminosity relation of quasars is
shown in Peterson et al.\ (2004) to be \footnote{All logarithms are base 10.}:

\begin{eqnarray}
\lefteqn{\mbox{log(\mbh}/10^8\mbox{ \msol}) = -0.12 (\pm 0.17) + } \nonumber \hspace{30pt} \\ 
  & & 0.79 (\pm 0.09) \mbox{ log}(\lambda L_\lambda (5100{\mbox{~\AA}})/10^{44}
\mbox{ ergs}^{-1}).
\end {eqnarray}

For objects which we do not have spectral information to determine accurate
virial BH masses, we can obtain a more crude estimate through this
\mbh-L$_{QSO}$ relation.  We also point out that Eq.\ 3 is based on
measurements of nearby AGNs, and currently there is little reason to believe
that the same normalization would apply to luminous quasars at $z=2$ where the
distribution of luminosities, Eddington ratio, and AGN lifetimes, may be
somewhat different (e.g. see Kollmeier et a. 2005).  The goal of this crude
method is thus only to gauge how a bigger scatter, uncertainties in the
overall BH mass normalization and small number statistics, can affect
interpretation.  And we emphasize that our final conclusions are based only on
virial \mbh\ estimates alone.

\subsection {Eddington Ratios of Quasars}

From \mbh\ estimates using the virial technique and assuming that the
bolometric luminosity $L_{\rm bol} \approx 9 \lambda
L_\lambda(5100\mbox{~\AA}$), as also assumed by Peterson et al. (2004) and
Kaspi et al. (2000), we can estimate the Eddington ratio, $\epsilon$, for 11
quasars at $z\approx 2$, and 4 at $z\approx 1$, which have published spectra
as a check for any unusual objects (e.g.  $\epsilon > 1$).  For $z\approx1$
quasars which lack the required $V$-band magnitude, we assume $B-V=0.3$, which
corresponds to a continuum of $f_\nu\propto\nu^{-0.5}$ (Vanden Berk et al.
2001).  Table 2, Col. 11, shows the Eddington ratio computed for 15 quasars
that have virial \mbh\ estimates.  All the quasars in our sample have $L_{\rm
bol} < L_{\rm Edd}$, with the only object, 4C~45.51, which is radio loud, near
the Eddington limit.  The Eddington ratio spans a wide range from
$0.02\le\epsilon\le 0.7$, with the top two highest efficiencies being RLQs and
the bottom two being RQQs.  This is probably due to a selection effect since
Kukula et al.~(2001) selected RLQ and RQQ quasars that are well matched in
optical luminosities, but required RLQs to be extremely radio luminous ($> 1$
Jy sources, or $\sim 3\times10^{28}$ to $\sim10^{29} \mbox{ W Hz}^{-1}$).
Because our sample size is small, and because a number of $z\approx2$ RLQs and
$z\approx1$ RQQs are missing virial \mbh\ measurements, the issue of a
correlation between radio loudness and BH mass and $\epsilon$, at a given
redshift, cannot be addressed here.  We only note that such claims of a
correlation are controversial even at low-$z$:  while several studies (e.g.
Pagani, Falomo, \& Treves 2003; Dunlop et al.  2003; and McLure \& Dunlop
2001) find that RLQs tend to have more massive \mbh\ and more powerful quasars
than RQQs at low $z$, several other studies find little to no correlation
(e.g. Ho 2002; Woo \& Urry 2002; Oshlack, Webster, \& Whiting 2002; Gu, Cao,
\& Jiang 2001).

\subsection {Measurement Uncertainties}

In this study, the robustness of the conclusions relies on understanding the
sources for both systematic and random errors.  In light of systematic errors,
the best way to minimize potential doubts is to apply all known systematic
errors (by offsetting the data) towards weakening the final conclusions.
Specifically, this can be done by using conversions that yield systematically
{\it small} \mbh\ and {\it bright} host luminosities.  We therefore take
measures to do so, as described below.  In this section we summarize the known
sources of random and systematic errors, some of which are collected from
previous discussions.

There are several important sources of {\it random} errors when trying to
interpret the scatter in the plots.  In particular, the literature measurement
of the emission-line widths are probably uncertain by $\sim 20-30\%$ due to
manual measurements and complications from absorption-line features and
probable Fe~{\sc ii} contamination.  This produces roughly a random scatter of
$\sim 0.2-0.3$ dex in \mbh, while lines which are absorbed in the wings may
produce underestimates.  The uncertainty in the \mbh\ measurement using the
virial technique has about 0.3-0.4 dex in scatter (factor of 2.5; Kaspi et al.
2000, Vestergaard 2002, and McLure \& Jarvis 2002, Peterson et al. 2004),
which is also random.  The uncertainties of the host galaxy magnitudes are
better quantified:  typically $\sim 0.2-0.75$ mag (Ridgway et al.  2001 and
Kukula et al. 2001).  $k$-correction introduces both a systematic bias (by
assuming the reddest possible SED) and a random component (the intrinsic SED
scatter of the hosts).  In the \mbh-L$_{QSO}$ method of estimating
\mbh, one uncertainty is in the intrinsic scatter in converting optical
luminosity, using $\lambda L_\lambda(5100\mbox{~\AA}$), into bolometric
luminosity.  This is effectively factored into the estimate of \mbh\ from
$\lambda L_\lambda(5100\mbox{~\AA}$), shown in Eq. 3.  As will be evident,
photometric errors (quasar and host galaxy luminosities) will not obfuscate
the conclusions, because they are small compared to the intrinsic scatter in
empirical relationships between \mbh-\mr, and the virial technique in
estimating \mbh.  And the results are robust against even these random sources
of error.

On the other hand, the robustness of the conclusion will depend entirely on
the treatment of {\it systematic} issues.  We address this issue by applying
known errors to the data in a way to yield a smaller \mbh\ and more luminous
host galaxies, thereby weakening the conclusions in the outcome.  For
instance, in computing the host galaxy \mr, the main systematic error comes
from the quasar removal.  In Ridgway et al.\ (2001) they find after running
simulations that the amount of over-subtraction of the host galaxy flux
depends on the host morphology.  Thus, we choose a morphology correction that
results in our hosts being typically 0.3--0.5 mag {\it brighter} than quoted
in their study.  In Kukula et al.\ (2001), they perform 2-dimensional image
decomposition, using a de Vaucouleurs model profile to extrapolate the host
both into the core and out to the wings.  Fortunately, assuming a de
Vaucouleurs profile often tends to {\it overestimate} host fluxes if the
intrinsic profile is shallower, e.g.  an exponential.  Furthermore, they hold
fixed the effective radii to 5--10 kpc in the fit -- large compared with even
typical quasar hosts at $z\gtrsim2$ (i.e.  measured through gravitational
lensing, Peng et al. in prep.), and Lyman break galaxies (Trujillo et al.
2005).  We thus feel confident about taking their host measurements at face
value to be reliable upper limits.  In converting to restframe $R$-magnitude,
there are uncertainties also in the $k$-correction:  we again decide to err on
the side of overestimating the host luminosity by assuming a red SED, typical
of present-day E/S0 galaxies.  The difference between a red SED and a typical
Lyman-break galaxy SED is only about 0.3 magnitudes in restframe $R$ out at
$z=2$, when observed in the $H$-band.  In addition, most of the NICMOS images
are not sufficiently deep to ascertain whether the hosts are bulge or disk
dominated; we attribute all the host galaxy light to the bulge, which is also
a conservative assumption.

The systematic errors for estimating \mbh\ are managed in the following way.
In particular, there is some uncertainty in the luminosity dependence
(expressed via the exponent, $\gamma$, in Eqs. 1 and 2) of the broad line
radius in using the virial mass estimators.  Depending on the line fitting
technique used, and which sub-sample is being studied (e.g.  high-luminosity
vs.  low-luminosity) $\gamma$ ranges from $0.5 \lesssim \gamma \lesssim 0.74$,
although a lower $\gamma$ is preferred in the newest calibrations for the UV
broad line regions (Kaspi et al. 2005).  For the C~{\sc iv} virial \mbh\
estimate, the new calibration uses a value of $\gamma = 0.53$ (M.~Vestergaard,
private communication), while $\gamma = 0.47$ for Mg~{\sc ii} (McLure \&
Jarvis 2002).  Also, in the virial \mbh\ estimate (\S~3.1) there is a
potential systematic error in the ``$f$-factor'' that normalizes the virial
\mbh\ to the \msig. This normalization of $f=5.5$, which increases \mbh\ by a
factor of 1.8 compared to Kaspi et al. (2000), is based on a modest sample
size of 16 objects (Onken et al. 2004).  Since this is the best determined
value currently we will utilize it.  However, leaving it out weakens the
result by 0.5 magnitude in \mr, which, by coincidence, corresponds closely to
results using the mass-luminosity relation to estimate \mbh.  Even this
measure would leave the qualitative conclusions unchanged, especially in light
of other steps we have taken to weaken the conclusions; we will present both
techniques side-by-side to emphasize this fact.  In estimating BH masses using
the \mbh-L$_{QSO}$ technique, Peterson et al.  (2004) discuss in more depth
the associated uncertainties.  We note that the Y-intercept, $-0.12$, of Eq.
3 has a significant uncertainty of $\pm0.17$, potentially causing a small
normalization difference in mass estimates between this and the virial
technique.  We also do not expect the normalization for low redshift AGNs be
identical to high redshift objects because of different selection, and perhaps
physical, issues.

There is one particular object, the RLQ 4C~45.51, which deviates significantly
from others and it is worthwhile to discuss the relevant uncertainties.  The
uncertainty in the BH mass is not known, because the uncertainty in measuring
the FWHM is not known.  If one assumes the FWHM is in error by 50\%, an
unlikely high error, the error in estimating \mbh\ is $\sim 0.4$ dex.  The
potential that the quasar is beamed adds an unknown systematic error into the
estimate of \mbh.  For the host galaxy, Kukula et al. (2001) quotes a generic
errorbar for the ensemble of 0.75 mag, barring small uncertainties in
$k$-correction.  All the known errors added constructively cannot explain its
large departure from the other data points.  Other indicators, such as the
high Eddington ratio ($\epsilon\sim 0.7$) and the extreme radio loudness
($\sim 3\times10^{28}$ to $\sim10^{29} \mbox{ W Hz}^{-1}$ from 151 MHz to 37
GHz, e.g. Wiren et al.  1992; Hales, Baldwin \& Warner 1993) point to the
possibility that 4C~45.51 may be an exceptional object.  Thus we will exclude
this object from our discussions and analysis below.

Lastly, our conclusions rest crucially on the applicability of local virial
calibrations to objects out at considerably higher redshift, $z=2$, and that
they apply for RLQs whose emission may be beamed towards us (Oshlack, Webster,
\& Whiting 2002; Greene \& Ho 2005c).  The propriety of applying these
calibrations to luminous quasars is more rigorously discussed, e.g.\ in Netzer
(2003) and Vestergaard (2004); we point out that the luminosity range of
quasars in this sample is spanned by those used to calibrate the virial
relationships locally.

In summary, to mitigate against systematic errors that may complicate the
interpretation, we have added known systematic offset corrections to the
data that would most weaken the conclusions.  Therefore, any remaining
deviation that we subsequently observe might be a lower bound on the true
difference if all the sources of systematic errors are identified as discussed
here.

\section {RESULTS}

\subsection {\mbh-\mr\ Relationship for $z>1$ Quasar Hosts}

Figure 1 shows the \mbh-\mr\ relationship derived for high-$z$ quasars (open
symbols) based on the virial mass measurements (Fig. 1{\it a}) or on the mass
versus quasar luminosity relation (Fig.  1{\it b}).  The round solid points
(``$z\approx 0$ norm. gal.'') are local calibrations of normal elliptical
galaxies obtained in various studies.  In particular, the \mbh\ measurements
for $z\approx 0$ sample are the revised values given in Tremaine et al.\
(2002), while their photometry is compiled from different sources and
reprocessed into $R$-band values by Bettoni et al.  (2003).  There are many
measurements of other galaxy types that can be (and have been) placed onto
this plot, such as radio galaxies (Bettoni et al.  2003) and low-$z$ quasar
and Seyfert hosts (e.g., McLure \& Dunlop 2002).  All of them essentially fall
along the solid line shown, and have comparable scatter (McLure \& Dunlop
2002; Bettoni et al.  2003).  The solid lines shown in Figures 1{\it a} and
1{\it b} are based on a fit to the normal galaxies data points (round solid)
performed by Bettoni et al.  (2003).  Converting their cosmology ($H_0 = 50$
km s$^{-1}$ Mpc$^{-1}$) into ours, the line has the following relation:

\begin {equation}
{\rm log}\,({\mathcal M}_{\rm BH}/{\mathcal M}_\odot) = -0.50 (\pm 0.06)M_R - 2.70 (\pm 1.35).
\end {equation}

\bigskip

\noindent {\it Consistency in the \mbh-\mr\ Relation Between $z\approx0$ and
$z\gtrsim2$.}\ \ \ \ \ Figures 1{\it a} and 1{\it b} show a comparison of the
techniques used to estimate BH masses out at high $z$, illustrating curiously
that most of the $z\gtrsim 2$ host galaxies already lie near the \mbh-\mr\
relationship at $z\approx 0$.  The $z\approx 1$ quasar hosts appear to lie
noticeably below the relation.  However, as we have not yet accounted for
luminosity fading, it is not yet obvious what is to be expected.  Most of the
attention should be focused on Figure 1{\it a} because the estimates of \mbh\
are more robust.  Figure 1{\it b}\ merely confirms the fact that even with a
larger scatter, and possibly a systematic normalization offset in BH mass, the
result does not change significantly.  Although the points using the crude
estimate agree qualitatively with the virial estimates there are some notable
exceptions that bring the average in Figure 1{\it b} considerably lower.  The
most discrepant points are the two extreme radio sources objects 4C~45.51 and
B2~2156+29 from Kukula et al. (2001), at least one of which has high Eddington
ratio.  In Figure 1{\it a}, the correlation seen for $z>1$ objects is probably
real and not due to strong selection effects in detecting the host galaxy,
despite well-known challenges in extracting host galaxies from underneath
luminous quasars.  While one does expect selection bias to hamper finding
hosts lying to the left of the sample points, this appears not to be a serious
problem because 23 host are detected out of 24 targets in Kukula et al.
(2001) and Ridgway et al.  (2001).  On the other hand, if host galaxies exist
to the right of the sample points, they would be {\it easier} to detect.
Thus, the fact that there is not a larger and a more uniform scatter to the
right suggests that the \mbh-\mr\ relation has already been established as
early as $z=2$, at least in RQQ hosts.

To further interpret Figures 1{\it a} and 1{\it b}, it is useful to keep in
mind that regardless of where the high-$z$ quasar host galaxies lie, by $z=0$
they must scatter around the solid line representing the relationship for
$z=0$ galaxies.  Thus, it is surprising that ignoring any luminosity
evolution, the \mbh-\mr\ relation of quasar host galaxies at $z\gtrsim 2$
already falls near the same relation as local galaxies.  To quantify the
similarity of $z\gtrsim2$ AGN hosts to the local relation we fit a line to the
RQQ objects (shown as dotted lines in Fig.~1), holding the slope fixed to the
value of $-0.5$ determined for low-$z$ galaxies (Equation 4).  As a caution we
leave RLQ objects out of the fit because of possible issues with beaming and
because of their higher Eddington ratios, as discussed in Section 3.3, even
though most objects do not deviate far from RQQs.  There is only one object
which is radio loud in Figure 1{\it a} (4C 45.51) and 4 objects in Figure
1{\it b} -- two of which have the highest host luminosity in the sample, and
are the farthest outliers.

In Figure 1{\it a}, the horizontal offset between the dotted and solid lines
is $-0.3$ mag when RLQ, 4C~45.51 is excluded.  If included, 4C~45.51 would
single-handedly shift the dotted line further to the right by 0.3 magnitudes,
even though it is only one out of 11 objects.  The conclusions are largely
immune to this shift, as evidenced later when evolution models are factored
in.  The average shift in Figure 1{\it a} is also somewhat biased by a single
RQQ, MZZ~9592 (green triangle near \mr$\approx -25$).  This host is the second
highest redshift object ($z=2.71$) in the sample, observed in restframe
$B$-band.  Compared to a blue powerlaw SED ($\nu^{-1}$) of coeval Lyman-break
galaxies, this object has $k$-correction value brighter by 1 magnitude (as
opposed to $\sim 0.3$ mag for other hosts).  In light of this and our use of
the E/S0 SED for $k$-correction, if the hosts are as blue as Lyman break
galaxies, the agreement between $z\approx 2$ and the local \mbh-\mr\ relation
would be even more striking, bringing the two closer by at least 0.3
magnitudes, bridging any difference between high $z$ bulges and $z=0$ normal
galaxies.

In Figure 1{\it b}, the offset in the dashed line is larger, $-0.6$ mag, which
indicates a normalization difference between the two estimates of \mbh.  This
line also excludes 4 out of 10 objects, which are RLQ hosts; including them
would shift the dashed line to the right by 0.2 magnitudes.

\bigskip

\noindent {\it \mbh/\mbulge\ Ratio at $z\approx 2$.}\ \ \ \ \ If the local
\mbh-\mr\ relation more fundamentally reflects a relationship between \mbh\
and \mbulge, rather than with the bulge luminosity, then the fact that
$z\gtrsim2$ hosts lie near the same relationship as $z\approx0$ data is
unexpected even before applying a correction for luminosity evolution.  About
half of $z\approx2$ hosts are nearly on the $z\approx0$ relation -- more, if
the SEDs of the hosts are bluer (hence fainter) than we have assumed.  If the
hosts starved their black holes (i.e. constant \mbh) and there has been no
galaxy merger since $z\approx2$, this implies that the bulge mass was lower at
high $z$: as the host faded, more stars would have to be formed in concert to
build up the bulge mass (luminosity), allowing the host galaxy to remain
roughly fixed in the \mbh-\mr\ diagram.  We can constrain the \mbh/\mbulge\
ratio at $z\approx 2$ if we assume an evolutionary scenario that the RQQ host
galaxies at $z\gtrsim 2$ are fully formed ellipticals.  The offset between the
dotted and solid lines (Figure 1{\it a}) allows a fading of $0.3$ mag on
average since $z\gtrsim2$ (less, if \mbh\ also increases or if the bulge SED
is blue).  Such a small amount of dimming is perhaps surprising, and we will
discuss several possible evolutionary scenarios below.  We show later that
under the premise that the hosts evolve like E/S0 galaxies, simple evolution
models predict fading by factors 1--2 magnitudes between $z=2$ and $z=0$
(Figure 2).  In this scenario, Figure 1 then indicates that the
\mbh/\mbulge\ ratio at $z=2$ is higher than locally by a factor of 3--6.  Rix
et al. (1999, 2001), through studying the hosts of gravitationally lensed
quasars, first noted that unless all the quasars are radiating near ($\epsilon
\gtrsim 0.5$) the Eddington limit, the \mbh/\mbulge\ ratio must be higher at
high-$z$ than today.  Ridgway et al. (2001) also came to this same conclusion
using the same assumption.  Here, the argument is made more stringent because
the black hole mass is measured simultaneously with the host galaxy light,
and the fact that \mbh\ can only grow with age.

\bigskip

\noindent {\it Deviations from the \mbh-\mr\ Relation.}\ \ \ \ \ The largest
departure from the dotted line in Figure 1{\it a} is 4C~45.51 at $z=2$, and
the two largest in Figure 1{\it b} are also radio loud quasar hosts, one of
which is 4C~45.51.  Taken at face value, its deviation is so large that the
apparent correlation seen for most $z\approx2$ hosts in Figure~1{\it a} may
not be clear-cut for all objects, if beaming is not an important factor to
consider when estimating \mbh.  However, it is important to keep in mind that
this may have nothing to do with the issue of RLQ vs.\ RQQ hosts specifically.
Rather, it could be a statement about {\it massive vs.\ less massive galaxies}
in general.  The object 4C~45.51 is a radio source whose extreme radio
emissions would likely require an unusually massive BH to produce.  If the
\mbh-\mbulge\ relation continues to hold at high-$z$, then a selection on
extreme \mbh\ (due to selecting on powerful radio sources) would then more
efficiently pick out massive hosts, even if RLQ and RQQs are hosted by the
same underlying parent distribution of galaxies.  The location of 4C~45.51 in
Figure 1{\it a}, especially, may show that {\it massive, thus rare} galaxies
may be closer to their end-stage of evolution at $z\approx2$ than less massive
ones.

On the other hand extreme radio loudness can itself indicate beaming, which
might cause \mbh\ to be under-estimated in RLQs.  If the RLQ emission is in
fact beamed, the continuum luminosity alone would lead to an overestimate of
\mbh\ by a factor proportional to $\sim \gamma_L$, where $\gamma_L$ now refers
to the Lorentz factor, and a proportionality constant that depends on the
viewing angle.  For 4C~45.51, it may be that $\gamma_L\gtrsim 1$ because of
its near-Eddington luminosity.  On the other hand, if the kinematics of the
BLR are disc-like and the axis of the jet is aligned with the axis of the
disc, we expect the FWHM of the broad emission lines in beamed objects to
appear systematically narrower than their intrinsic width (Wills \& Browne
1986).  Since \mbh\ scales as FWHM$^2$, the inclination effect may counteract
the Doppler boosting.  Thus, the net effect on measuring \mbh\ depends on
better knowing the BLR inclination and the intrinsic $\gamma_L$.  For RLQs
with low inclination and low Doppler boosting, we probably underestimate \mbh.
However, it is difficult to draw a strong conclusion without a larger sample
of objects.

Finally, as we already pointed out but worth emphasizing, the horizontal shift
in \mr\ of the points in Figure 1 depends rather little on the E/S0 SED we
assumed, and uncertainty about the potential of a disk component in the host.
If we had assumed either a bluer SED, or assume that some fraction of the
bulge light belongs to the disk, they would both necessarily shift the points
to the left, further strengthening the conclusion that hosts at $z=2$ 
already lie close to the local \mbh-\mr\ relation.

\subsection {Host Galaxy Luminosity Evolution}

The near coincidence in the \mbh-\mr\ relation between $z\gtrsim 2$ RQQ hosts
and the present epoch must reflect several compensating effects, as time has
not stood still since $z\sim 2$.  Therefore, it places interesting constraints
how the host galaxies can evolve since the Universe was 2-3 Gyr old.  At low
$z$ ($z\lesssim1$), McLeod \& Rieke (1995) and Dunlop et al.~(2003) show that
most very luminous quasars ($M_V < -23.5$ mag) live in ellipticals, with the
variety of host galaxies increasing towards lower-luminosity AGNs.  At
intermediate redshifts ($1\lesssim z \lesssim 1.5$), several studies (Kukula
et al. 2001, S\'anchez \& Gonz\'alez-Serrano 2003) also find that the hosts
may be preferentially early-type.  At $z>1.5$, host galaxies are poorly
resolved, but there is circumstantial evidence to suggest that they are also
primarily ellipticals, fully formed.  Hence, we test the quite natural
null-hypothesis that the quasar host galaxies had fully formed stellar bodies
and have merely faded since $z\sim 2$.  Alternatively, the hypothesis can be
phrased as:  was the ratio of \mbh\ to \mbulge\ the same at all earlier epochs
than observed now?

The 0.3 mag displacement of the dotted line from the solid line in Figure
1{\it a} implies that host galaxies at a given \mbh\ appeared brighter in the
past only by $\left<|dm/dz|\right> \approx 0.20$ mag in $R$.  If host galaxies
did only fade since $z\sim2$, and by more than $\left<|dm/dz|\right>= 0.20$
mag, they would now appear displaced off the $z=0$ relation (solid line).  If
the host galaxies at $z=2$ have bluer SED than assumed in our $k$-corrections,
the maximal fading rate would be even slower.  For example, if we use the SED
of an Scd-type galaxy in our $k$-correction, the fading rate would be
consistent with essentially $\left<|dm/dz|\right> \approx 0$ mag -- i.e. no
luminosity fading.  Of course, if they are to gain in \mbh, through accretion
but not grow in \mbulge, the allowed fading rate for consistency with $z\sim
0$ would be smaller still.

To interpret the fading rate in terms of the host stellar population content,
we generate a grid of stellar synthesis models from Bruzual \& Charlot (2003)
with a Salpeter initial stellar mass function.  The two sets of models (Fig.
2) have formation redshift $z_f= \{2, 5\}$, i.e. universe age = \{3.2, 1.2\}
Gyr, each with star formation rates that decrease exponentially with a decay
time constant of $\tau = \{0, 0.5, 1, 2, 5\}$ Gyr, where $\tau=0$ corresponds
to an instantaneous burst followed by passive evolution.  For all the models
except that for $\tau=5$ Gyr, we find that $\left<|dm/dz|\right> \approx 0.9$
mag over the range $0 \le z \le 2$.  Figure 2{\it b} illustrates how the
fading scenarios translate into $B-R$ color of the host galaxies.  Only the
models with $\tau = \{0, 0.5, 1, 2\}$ Gyr evolve to early-type galaxy colors
by $z=0$; they have a fading rate $\left<|dm/dz|\right> \approx 0.9$ mag.  The
$\tau=5$ Gyr scenario results in a present day color 0.4--0.5 mag bluer than
average early-type galaxies.

If the $z=1... 3$ host galaxies were simply to fade (and do not gain in \mbh),
resulting in present-epoch early-type host galaxies we need to account for
this luminosity evolution, before comparing to the local \mbh-\mr\ relation.
We do this in Figure 3 by taking the measurements of distant hosts (open
symbols) from Figure 1 and simply shift the open points horizontally by an
amount $|dm/dz|=0.8$ mag (see Figure 2), appropriate for a {\it passively}
fading red bulge.  Figures 3{\it a} and 3{\it b} show that in this scenario,
quasar hosts at $z\approx 1$ evolve onto the $z\approx 0$ relationship.  Some
of the deviations may be explained by our $k$-correction, or if the deviant
galaxies underwent a starburst at $z\approx 1$.  However, for host galaxies at
$z\approx2$, what is striking is that nearly all would evolve to be fainter
than local galaxies with similar mass BHs.  The average offset of $z\approx2$
RQQ hosts after such fading would now be at least 1.3 mag, and as high as
$\sim 2$ mag, below the local \mbh-\mr\ relation.  Hence our data are {\it
inconsistent} with simple evolutionary or passively fading models for an
early-type galaxy.  In fact, the only model that is marginally consistent with
the present-day luminosity constraint, $|dm/dz| \lesssim 0.20$ mag, requires
an evolution that has a slow decline in star formation rate of $\tau=5$ Gyr,
and that evolves to a blue galaxy ($B-R=1.1$ mag) by $z=0$ (Fig.  2{\it b}).
This is only possible if all descendants of $z\sim 2$ RQQ hosts were blue
bulges now.  Even the most extreme case of an elliptical, fully formed by $z_f
= \infty$ and passively fading, would have $|dm/dz| \approx 0.6$ (van Dokkum
\& Franx 2001) -- still three times larger than allowed for consistency with
no mass evolution.  On the other hand, for largely passive evolution of the
host galaxy populations (with $z_f\le5$), the only way to reconcile Figure 3
with the local \mbh-\mr\ relationship is to require the hosts to undergo
mergers that increase the galaxy stellar mass by $\approx 3\times$ (1.3
magnitude).  However, in this process, they must not gain an equal proportion
of \mbh\ at the merger, which only takes them parallel to the solid line.  Our
conclusions that $z\sim 2$ hosts appear too faint to be reconciled easily with
present-day counter-parts of the same \mbh\ would be even stronger for any
other plausible $k$-correction one could make.

In the discussion above we have not considered that \mbh\ may also evolve.  We
collect here a few other interesting implications gleaned from Figures 1 and
3:

\begin {itemize}

\item If evolution occurs on the two-dimensional \mbh-\mr\ plane, the curious
fact that $z=2$ hosts already lie close to the $z=0$ relation implies that
evolution happens mostly {\it along} the relation, allowing for a modest
amount of fading.  Black hole growth, fading and the acquisition of new stars
through mergers or star-formation seem to be balanced.

\item The BH masses for most of the host galaxies in our study at $z\gtrsim1$
are already higher than that expected for $L_*$ galaxies ($M_R^* = -20.88$,
Brown et al.  2001).  Therefore, regardless of the stellar population of the
$z=2$ host galaxies brighter than \mr $< -23$ mag (as in the Kukula et al.
2001 and Ridgway et al.  2001 sample), they cannot evolve to $\lesssim L_*$
galaxies today and yet agree with the local \mbh-\mr\ relation.

\item While the hosts at $z\gtrsim2$ may subsequently increase in luminosity
moderately via merging, and to increase \mbh, Figure 1 shows that the most
massive BHs may not grow by 3 to 6 times without becoming more massive than
the biggest BHs seen at low $z$.  If the bulge mass grows proportionally with
\mbh, this would also suggest that the host galaxies do not increase in mass
more than a factor of 3 to 6.

\item Figure 1{\it b} shows that the host galaxies of $z=1$ quasars are
roughly as luminous as the hosts at $z\gtrsim2$ for objects that have
comparable \mbh; Figure 1{\it a} is somewhat equivocal on this point because
of the small number of points, despite the smaller scatter than in Figure
1{\it b}.  The correspondence may be fortuitous.  Our evolutionary scenario
suggests that the $z\approx1$ hosts are consistent with a simple fading
scenario for an E/S0 galaxy with $z_f=5$, while the evolution of $z\gtrsim2$
hosts may be more complex.

\item Because \mbh/\mbulge\ ratio is higher in the past than today, this might
suggest that the \msig\ relation has a different zeropoint earlier in time.
Furthermore, if the dominant route of galaxy evolution since $z\approx2$ is by
way of galaxy merging rather than {\it in situ} star formation, it would
suggest that the \msig\ at high redshift has a {\it steeper} slope than that
observed today (Gebhardt et al. 2000, Ferrarese \& Merritt 2000).  A steeper
slope in the \mbh\ vs. $\sigma$\ plot might then allow galaxies (as they lie
in Fig. 3) to grow in mass through mergers at a faster rate than the black
hole mass, without evolving parallel to the local \mbh-\mr\ relation.

\end {itemize}

\section {DISCUSSION AND CONCLUSION}

In this study, we have tested the hypothesis that both the stellar bodies of
the host galaxies and the central black holes were fully formed by $z_f=5$,
evolving secularly thereafter.  This is motivated by the evidence that $z\le1$
hosts of luminous quasars live in elliptical galaxies (e.g.  McLeod \& Rieke
1995, and Dunlop et al. 2003).  Moreover, several studies (e.g. Kukula et al.
2001, S\'anchez \& Gonz\'alez-Serrano) argue that host galaxies as early as
$z\gtrsim2$ may be fully formed, early-types, undergoing passive fading.  Our
analysis of the existing data has shown that the \mbh-\mr\ relation of
high-$z$ quasar hosts is nearly identical to that of low-$z$ galaxies (Figure
1).

We have mostly focused our attention on $z\approx 2$ objects.  On the other
hand, $z\approx1$ hosts do appear to deviate from the local \mbh-\mr\
relationships in Figure 1, which suggests they may be fully consistent with a
secularly evolving E/S0 model with $z_f=5$.  However, currently, this evidence
is limited by small number statistics.

The close agreement between $z\gtrsim2$ and $z\approx0$ relation is surprising
given that the underlying correlation seen at $z=0$ is fundamentally one
between \mbh\ and \mbulge, and more superficially between \mbh\ and bulge
luminosity.  There are several ramifications of this finding, one of which is
that the bulge luminosity does not evolve as much as simple stellar synthesis
models would predict, thus allowing the bulge and \mbh\ to do so in near
lock-step upwards.  We have shown, using a set of conservative assumptions
about the stellar population of the host, that the stellar bulge mass,
\mbulge, at a given \mbh, is probably lower in the past than today:  If there
is luminosity evolution, but no \mbh\ evolution, then the \mbh/\mbulge\ ratio
is larger at $z>1$ by almost the same factor that the galaxy dims due to
passive evolution.  Because fading (passive or otherwise) would cause the host
galaxies at $z\approx2$ to significantly over-shoot the $z\approx0$ relation,
there has to be a build up in \mbulge\ stellar mass towards $z=0$, whether by
star formation or by mergers (without increasing \mbh\ in the same
proportion).  This was previously suggested by Rix et al. (1999, 2001) and
Ridgway et al.\ (2001).

As BHs only increase in mass, Figure 1{\it a} puts a strong constraint on how
much the host galaxies can fade to arrive at the local \mbh-\mr\ relation by
$z=0$, regardless of their stellar population content at $z\gtrsim1$. Luminous
host galaxies (\mr$\lesssim-23$ mag) with luminous quasars cannot fade to
become $L_*$ galaxies today, unless the mass estimators calibrated locally
systematically and significantly overestimate \mbh\ at high $z$.

While this study supplies more questions than answers regarding the nature of
galaxy-BH evolution, we show that the \mbh-\mr\ diagram provides very useful
constraints on the evolutionary paths of galaxies.  Thus, it may be worthwhile
to extend this diagnostic by establishing locally the empirical relationship
between \mbh\ and bulge luminosity in multiple filters, since galaxies of
different morphological types are expected to trace distinct paths in this
space.  In this parameter space of \mbh\ vs. multi-color bulge luminosity
diagram the locus of points for high-$z$ galaxies can be used to predict the
directions in which galaxies evolve.  Galaxies in this diagram are not allowed
to arbitrarily roam in color and luminosity because their evolutionary paths
must ultimately take them to the local \mbh-bulge luminosity relationship.  To
understand how exactly galaxies traverse the phase diagram, it would be useful
to obtain deep imaging of the environments of high-$z$ host galaxies in order
to constrain their merger rates.

Thus far, we have not discussed how the presence of dust may affect our
conclusions.  Sub-mm observations of $z\gtrsim 2$ quasars (Isaak et al.  2002;
Bertoldi et al.  2003; Knudsen, van~der~Werf, \& Jaffe 2003) indicate that at
least some host galaxies appear to be highly star forming and hence presumably
highly obscured.  If this were generically true of all high-$z$ hosts, what we
observe in the rest-frame $V$-band might be the dim light that filtered
through a dense screen.  In later phases, star formation may use up gas and
dust, while supernovae explosions may blow holes in the ISM, causing host
galaxies to become more transparent with time.  If this were to happen, the
net effect of the fading populations and the decreasing optical depth on the
absolute luminosity of the host galaxies is not clear.  It is conceivable that
the host does not evolve much in optical luminosity, if the galaxy dust
extinction dropped at the same rate as fading of the stars.  The monkey-wrench
thrown into the interpretation by the presumption of dust is yet another
reason why it would be worthwhile to establish a multi-color vs.
\mbh\ ``phase'' diagram, especially towards the rest-frame infrared.  However,
despite the possibility of dust, its presence must somehow conspire with star
formation to preserve the good correlation seen at high-$z$ (Figure 1), over 4
magnitudes in host luminosity.  This suggests that either the distribution of
dust is regular across galaxy-types, or that dust obscuration is not too
significant in a number of galaxies by $z\approx2$.

In computing the $k$-correction, we assumed an SED appropriate for a {\it
current-day} E/S0 galaxy, but galaxies at $z\approx2$ are most likely to be
bluer.  Therefore the hosts are probably fainter in the rest-frame $R$-band
than what we computed, and the \mbh-\mr\ relationship for $z=2$ hosts would be
even more indistinguishable from $z\approx0$ normal galaxies.  This conclusion
relies heavily on the assumption that the techniques used to estimate \mbh\ at
low $z$ do not systematically and significantly overestimate \mbh\ at high
$z$.  Currently, all \mbh\ measurements are tied to locally calibrated values.
It remains unproven that the AGN broad-line regions at high $z$ have similar
structure at low $z$.  Because understanding the link between \mbh\ and galaxy
bulges is fundamentally important to a coherent knowledge of galaxy evolution
in general, it is important to more fully understand how locally calibrated
\mbh\ measurement techniques apply to high $z$.

Lastly, we prelude in the passing that a significantly larger sample of quasar
host galaxies at $z \gtrsim 2$ from gravitational lensing has been analyzed to
study the \mbh-\mr\ relationship.  The results from that study further
strengthen the conclusions presented here (Peng et al. in preparation).

\bigskip

\bigskip

\noindent We thank Daniel Eisenstein, Ann Zabludoff, and Dennis Zaritzky,
Marianne Vestergaard, Roeland van der Marel, Swara Ravindranath, and Masami
Ouchi, for discussions and advice.  We would also like to thank the anonymous
referee for useful suggestions and comments.  The work of CYP was performed in
part under contract with the Jet Propulsion Laboratory (JPL) funded by NASA
through the Michelson Fellowship Program.  JPL is managed for NASA by the
California Institute of Technology.  CYP is also grateful to STScI for support
through the Institute Fellowship Program.  LCH is supported by the Carnegie
Institution of Washington and by NASA grants from the Space Telescope Science
Institute (operated by AURA, Inc., under NASA contract NAS5-26555).  Support
for EJB was provided by NASA through Hubble Fellowship grant
\#HST-HF-01135.01 awarded by the Space Telescope Science Institute, which is
operated by the Association of Universities for Research in Astronomy, Inc.,
for NASA, under contract NAS 5-26555.  This research has made use of the
NASA/IPAC Extragalactic Database (NED) which is operated by the Jet Propulsion
Laboratory, California Institute of Technology, under contract with the
National Aeronautics and Space Administration.

\newpage

\begin {references}

\reference{} 
Akritas, M. G., \& Bershady, M. A. 1996, \apj, 470, 706

\reference{} 
Aldcroft, T. L, Bechtold, J., \&  Elvis, M. 1994, \apjs, 93, 1

\reference{} 
Aretxaga, I., Terlevich, R. J., \& Boyle, B. J. 1998, \mnras, 296, 643

\reference{} 
Barth, A. 2004, in Carnegie Observatories Astrophysics Series, Vol.  1:
Coevolution of Black Holes and Galaxies, ed. L. C. Ho (Cambridge: Cambridge
Univ. Press), in press

\reference{} 
Barth, A.~J., Greene, J. E., \& Ho, L.~C. 2005, \apj, 619, L151

\reference {}
Bertoldi, F., Carilli, C. L., Cox, P., Fan, X., Strauss, M. A., Beeleen, A.,
Beelen, A., \& Zylka, R. 2003, \aap, 406, 55

\reference {}
Bettoni, D., Falomo, R., Fasano, G., \& Govoni, F. 2003, \aap, 399, 869

\reference {}
Blandford, R. D., \& McKee, C. F. 1982, \apj, 255, 419

\reference{}
Boyle, B. J., Fong, R., Shanks, T., \& Peterson, B. A. 1990, \mnras, 243, 1

\reference{} 
Brown, W. R., Geller, M. J., Fabricant, D. G., \& Kurtz, M.  J.  2001, \aj,
122, 714

\reference{}
Bruzual, G., \& Charlot, S. 2003, \mnras, 344, 1000

\reference{}
Carollo, C. M. 1999, \apj, 523, 566

\reference{}
Coleman, G. D., Wu, C.-C., \& Weedman, D. W. 1980, \apjs, 43, 393

\reference{} 
Colina, L., Bohlin, R., \& Castelli, F. 1996, Instrument Science Report
CAL/SCS-008


\reference{}
Dunlop, J. S., McLure, R. J., Kukula, M. J., Baum, S. A., O'Dea, C. P., \&
Hughes, D. H. 2003, \mnras, 340, 1095


\reference{}
Erwin, P., Graham, A.~W., \& Caon, N. 2002, astro-ph/0212335

\reference{}
Falomo, R., Kotilainen, J., \& Treves, A. 2001, \apj, 547, 124

\reference{}
Ferrarese, L., \& Merritt, D. 2000, \apjl, 539, L9

\reference{}
Ferrarese, L., Pogge, R. W., Peterson, B. M., Merritt, D., Wandel, A., \&
Joseph, C. L. 2001, \apj, 555, 79

\reference{} 
Fynbo, J. U., Burud, I., \& M{\o}ller, P. 2002, NewAR, 46, 193

\reference{} 
Gebhardt, K., et al. 2000a, \apj, 539, L13

\reference{} 
------. 2000b, \apj, 543, L5

\reference{} 
Greene, J.~E., \& Ho, L.~C. 2005a, \apj\ submitted

\reference{} 
------. 2005b, \apj\ submitted

\reference{} 
------. 2005c, \apj, 630, 122 

\reference{}
Gu, M., Cao, X., \& Jiang, D. R. 2001, \mnras, 327, 1111

\reference{}
H\"aring, N., \& Rix, H.-W. 2004, \apjl, 604, 89

\reference{}
Hales, S. E. G., Baldwin, J. E., \& Warner, P. J., 2003, \mnras, 263, 25

\reference{}
Ho, L.~C. 1999, in Observational Evidence for Black Holes in the Universe,
ed. S. K. Chakrabarti (Dordrecht: Kluwer), 157

\reference{}
Ho, L.~C. 2002, \apj, 564, 120

\reference{} 
Hogg, D. W., Baldry, I. K., Blanton, M. R., \& Eisenstein, D. J.  2004,
astro-ph/0210394

\reference{}
Hutchings, J. B. 2003, \aj, 125, 1053

\reference{}
Hutchings, J.~B., Frenette, D., Hanisch, R., Mo, J., Dumont, P.~J., Redding,
D.~C., \& Neff, S.~G. 2002, \aj, 123, 2936

\reference{} 
Isaak, K. G., Priddey, R. S., McMahon, R. G., Beelen, A., Peroux, C., Sharp, 
R. G., \& Withington, S. 2002, \mnras, 329, 149

\reference{} 
Ivanov, V. D., \& Alonso-Herrero, A. 2003, Ap\&SS, 284, 565

\reference{} 
Jackson, N., \& Browne, I. W. A. 1991, \mnras, 250, 414

\reference{} 
Jauncey, D. L., Batty, M. J., Wright, A. E., Peterson, B. A., \& Savage,
A. 1984, \apj, 286, 498

\reference{}
Kaspi, S., Smith, P.~S., Netzer, H., Maoz, D., Jannuzi, B.~T., \& Giveon,
U. 2000, ApJ, 533, 631

\reference{} 
Kaspi, S., Maoz, D., Netzer, H., Peterson, B. M., Vestergaard,
M., \& Jannuzi, B. T. 2005, \apj, 629, 61

\reference{}
Knudsen, K. K., van der Werf, P. P., \& Jaffe, W. 2003, \aap, 411, 343

\reference{} 
Kochanek, C. S., Falco, E. E., Impey, C. D., Leh\'ar, J., McLeod, B. A., \&
Rix, H.-W. 1999, in After the Dark Ages: When Galaxies were Young (the 
Universe at 2 $\le z \le$ 5), ed. S. Holt \& E. Smith (New York: AIP), 163

\reference{} 
Kollmeier, J. A., et al. 2005, astro-ph/0508657

\reference{} 
Kormendy, J. 2004, in Carnegie Observatories Astrophysics Series, Vol.  1:
Coevolution of Black Holes and Galaxies, ed. L. C. Ho (Cambridge: Cambridge
Univ. Press), in press.

\reference{}
Kormendy, J., \& Gebhardt, K. 2001, in 20th Texas Symposium on Relativistic 
Astrophysics, ed. H. Martel \& J.~C. Wheeler (Melville: AIP), 363

\reference{}
Kormendy, J., \& Richstone, D. 1995, \araa, 33, 581


\reference{} 
Kuhlbrodt, B., Orndahl, E., Wisotzki, L., \& Jahnke, K. 2005, astro-ph/0503284

\reference{}
Kukula, M., Dunlop, J.~S., McLure, R. J., Miller, L., Percival, W.~J.,
Baum, S. A., \& O'Dea, C.~P. 2001, \mnras, 326, 1533

\reference{}
Lacy, M., Gates, E. L., Ridgway, S. E., de Vries, W., Canalizo, G., Lloyd, J.
P., \& Graham, J. R. 2002, \aj, 124, 3023

\reference{}
Laor, A. 2001, \apj, 553, 677

\reference{} 
Lehnert, M. D., van Breugel, W. J. M., Heckman, T. M., \& Miley, G. K. 1999,
\apjs, 124, 11

\reference{}
Magorrian, J., et al.  1998, \aj, 115, 2285

\reference{} 
Marconi, A., \& Hunt, L. K. 2003, \apj, 589, L21

\reference{}
McLeod, K. K., \& Rieke G. H. 1995, \apjl, 454, L77


\reference{}
McLure, R. J., \& Dunlop, J. S. 2002, \mnras, 331, 795

\reference{}
McLure, R. J., \& Jarvis, M. J. 2002, \mnras, 337, 109

\reference{}
Merritt, D., \& Ferrarese, L. 2001, \mnras, 320, L30

\reference{}
Nelson, C.~H., Green, R. F., Bower, G., Gebhardt, K., \& Weistrop, D. 2004,
\apj, 615, 652

\reference{} 
Netzer, H. 2003, \apj, 583, L5

\reference{} 
Onken, C. A., et al. 2004, \apj, 615, 645

\reference{} 
Oshlack, A. Y. K. N., Webster, R. L., \& Whiting, M. T. 2002, \apj, 576, 81

\reference{} 
Pagani, C., Falomo, R., \& Treves, A. 2003, \apj, 596, 830

\reference{} 
Peng, C. Y., et al.  2005, \apj\ submitted

\reference{} 
Peterson, B. M. 1993, \pasp, 105, 247

\reference{} 
Peterson, B. M. et al. 2004, \apj, 613, 682

\reference{} 
Press, W.~H., Teukolsky, S.~A., Vetterling, W.~T., \& Flannery, B.~P. 1992, 
Numerical Recipes in Fortran (2nd Ed.; Cambridge: Cambridge Univ. Press) 

\reference{} 
Ridgway, S.~E., Heckman, T.~M., Calzetti D., \& Lehnert, M. 2001, \apj, 550,
122

\reference{} 
Rix, H.-W., Falco, E. E., Impey, C. D., Kochanek, C. S., Leh\'ar, J., McLeod,
B.  A., Munoz, J., \& Peng, C. Y. 1999, astro-ph/9910190

\reference{} 
Rix, H.-W., Falco, E. E., Impey, C. D., Kochanek, C. S., Leh\'ar, J., McLeod,
B.  A., Munoz, J., \& Peng, C. Y. 2001, ASP Conf.~Ser.~237: Gravitational
Lensing: Recent Progress and Future Go, 169

\reference{}
S\'anchez, S. F., \& Gonz\'alez-Serrano, J. I. 2003, \aap, 406, 435

\reference{}
Schlegel, D.~J., Finkbeiner, D.~P., \& Davis, M. 1998, \apj, 500, 525

\reference{} 
Stickel, M., \& K\"uhr, H. 1993, \aaps, 101, 521

\reference{} 
Tremaine, S., et al. 2002, \apj, 574, 740

\reference{} 
Trujillo, I., et al. 2005, \apj\ submitted, astro-ph/0504225

\reference{}
Vanden Berk, D. E., et al. 2001, \aj, 122, 549

\reference{}
van Dokkum, P. G., \& Franx, M. 2001, \apj, 553, 90

\reference{}
V\'eron-Cetty, M.-P., \& V\'eron, P. 1996, ESO Sci. Rep. No. 17. ESO
Publications, Garching

\reference{}
Vestergaard, M., \& Wilkes, B. J. 2001, \apjs, 134, 1

\reference{}
Vestergaard, M. 2002, \apj, 571, 733

\reference{}
Vestergaard, M. 2004, \apj, 601, 676

\reference{}
Vestergaard, M., et al. 2005, \apj\ submitted

\reference{} 
Wandel, A., Peterson, B.~M., \& Malkan, M.~A. 1999, \apj, 526, 579

\reference{} 
Wills, B.~J., \& Browne, I.~W.~A. 1986, \apj, 302, 56

\reference{} 
Wiren, S. et al. 1992, \aj, 104, 1009


\reference{} 
Woo, J.-H., \& Urry, C. M. 2002, \apjl, 581, L5


\reference{} 
Zitelli, V., Mignoli, M., Zamorani, G., Marano, B., \& Boyle, B.  J. 1992,
\mnras, 256, 349

\end{references}

\clearpage


\begin{deluxetable}{|ll|rcccc|ccc|l|}
\tabletypesize{\scriptsize}
\tablewidth{0pt}
\tablecaption {Quasar and Host Galaxy Data Compiled from Literature}
\tablehead{Object & $z$ & Filter & \multicolumn{4}{c|} {NICMOS Mag} & \multicolumn{2}{|c}{Quasar} & Radio  & References/Comments\\
       &          &        & Host  &  Err  &  Quasar  &  Err &    Filter  & Mag  & Loud? & \\
    (1)         &   (2)  &  (3)   & (4)    & (5) &  (6)  & (7) & (8)   & (9) & (10) & (11)}
\startdata
SGP5:46     	& 0.955  & F110M  & 20.09  & 0.4 & 19.45 & 0.3 & $V^*$ & 19.7 & N  & 1, assumed quasar ($B-V$)=0.3 \\
BVF225      	& 0.910  & F110M  & 20.10  & 0.4 & 17.90 & 0.3 & $V^*$ & 19.2 & N  & 1 \\
BVF247      	& 0.890  & F110M  & 18.87  & 0.4 & 20.13 & 0.3 & $V^*$ & 19.5 & N  & 1 \\
BVF262      	& 0.970  & F110M  & 19.85  & 0.4 & 19.24 & 0.3 & $V^*$ & 19.4 & N  & 1 \\
PKS 0440$-$00  	& 0.844  & F110M  & 18.79  & 0.4 & 18.42 & 0.3 & $V^*$ & 19.1 & Y  & 1 \\
PKS 0938+18	& 0.943  & F110M  & 19.46  & 0.4 & 19.78 & 0.3 & $V$ & 18.9 & Y  & 1, assumed quasar ($B-V$)=0.3 \\
3C~422		& 0.942  & F110M  & 18.24  & 0.4 & 17.85 & 0.3 & $V$ & 18.9 & Y  & 1, assumed quasar ($B-V$)=0.3 \\
MRC~2112+172	& 0.878  & F110M  & 18.06  & 0.4 & 18.85 & 0.3 & $V$ & 17.8 & Y  & 1, assumed quasar ($B-V$)=0.3 \\
4C~02.54	& 0.976  & F110M  & 19.28  & 0.4 & 17.57 & 0.3 & $V$ & 18.5 & Y  & 1, assumed quasar ($B-V$)=0.3 \\
\tableline
SGP2:11		& 1.976  & F165M  & 20.64  & 0.75 & 18.96 & 0.3 & $B^*$ & 20.9 & N  & 1 \\
SGP2:25		& 1.868  & F165M  & 19.88  & 0.75 & 19.59 & 0.3 & $B^*$ & 20.7 & N  & 1 \\
SGP2:36		& 1.756  & F165M  & 19.73  & 0.75 & 19.97 & 0.3 & $B^*$ & 20.7 & N  & 1 \\
SGP3:39		& 1.964  & F165M  & 19.75  & 0.75 & 19.53 & 0.3 & $B^*$ & 20.8 & N  & 1 \\
SGP4:39		& 1.716  & F165M  & 21.59  & 0.75 & 18.85 & 0.3 & $B^*$ & 20.8 & N  & 1 \\
PKS 1524$-$13	& 1.687  & F165M  & 19.29  & 0.75 & 18.03 & 0.3 & $B$ & 20.0 & Y  & 1 \\
B2$-$2156+29	& 1.753  & F165M  & 17.81  & 0.75 & 17.91 & 0.3 & $B$ & 19.7 & Y  & 1 \\
PKS 2204$-$20	& 1.923  & F165M  & 20.63  & 0.75 & 18.54 & 0.3 & $B$ & 20.1 & Y  & 1 \\
4C~45.51	& 1.992  & F165M  & 17.79  & 0.75 & 17.41 & 0.3 & $B$ & 20.1 & Y  & 1 \\
\tableline
MZZ 9744    	& 2.735  & F160W  & 21.73  & 0.5 & 20.02 & 0.05 & $B^*$ & 21.4 & N & 2 \\
MZZ 9592    	& 2.710  & F160W  & 20.70  & 0.1 & 19.57 & 0.03 & $B^*$ & 21.8 & N & 2 \\
MZZ 1558    	& 1.829  & F165M  & 20.64  & 0.2 & 19.08 & 0.04 & $B^*$ & 21.5 & N & 2 \\
MZZ 11408   	& 1.735  & F165M  & 20.78  & 0.4 & 21.08 & 0.06 & $B^*$ & 21.9 & N & 2 \\
MZZ 4935    	& 1.876  & F165M  & 22.00  & 0.4 & 21.23 & 0.06 & $B^*$ & 21.8 & N & 2 \\
\tableline
CTQ 414         & 1.29   & F160W  & 19.67  & 0.1 & 18.09 & 0.2  & $V$ & 19.85 & ? & 3 \\
\tableline
\enddata 
\tablecomments {
Col. (1): Object name.
Col. (2): Redshift.
Col. (3): {\it HST} Filter.
Col. (4-7): Apparent magnitude and their published uncertainties, in the Vega
	    magnitude system, corrected for extinction from Schlegel et al.
	    (1998).
Col. (8/9): Quasar magnitude (corrected for extinction), in the Vega magnitude
	    system, corresponding to the filter in Col. (8).  Filters with
	    superscript $^*$ are photographic magnitudes.  The $B$ and
	    $V$-band magnitudes of the quasars in Kukula et al. (2001) sample
	    are from V\'eron-Cetty \& V\'eron (1996), and references therein,
	    while the $B$-band magnitudes for MZZ objects are from Zitelli et
	    al.  (1992).  Where $V$-band magnitude is needed and unavailable,
	    we used $(B-V)=0.3$, which corresponds to $f_\nu\propto\nu^{-0.5}$.
Col. (10): Radio-loud quasar or radio-quiet quasar.
Col. (11): The photometry for each set of objects comes from the references
          shown.  
References.--- (1) Kukula et al. 2001; (2) Ridgway et al. 2001; (3) Peng et al. 2005.
}
\end{deluxetable}


\begin{deluxetable}{|llr|cc|cc|ccc|c|}
\tabletypesize{\scriptsize}
\rotate
\tablewidth{0pt}
\tablecaption {Quasar and Host Galaxy Derived Quantities}
\tablehead{Object	     &$z$    &	DM    &  Host   & Quasar &    Quasar  &  \mbh	&   Emission & Observed & \mbh  &  $L_{\rm bol}/L_{\rm Edd}$ \\
	     		     &	      &  &	$M_R$  & $M_V$  & log~$\lambda L_\lambda({\mbox{5100~\AA}})$  &  & Line  & FWHM  &  & \\
                             &        & (mag) & (mag) & (mag) & (erg s$^{-1}$) &($10^9$ \msol) &        &    (\AA)     & ($10^9$ \msol) &\\
    (1)     &   (2)   &  (3)  &     (4)  &     (5)      &  (6)  &  (7) & (8)        & (9)     & (10) & (11)   }
\startdata
SGP5:46	    &	0.955 &	43.98 &	$-$23.00 & $-$23.18	& 44.67	& 0.26 & Mg~{\sc ii} &	55    &	0.16 &	0.21 \\
BVF225	    &	0.91  &	43.85 &	$-$22.81 & $-$24.51	& 45.21	& 0.68 & ---	    &	---   &	---  &	---  \\
BVF247	    &	0.89  &	43.79 &	$-$23.97 & $-$22.23	& 44.29	& 0.13 & --- 	    &	---   &	---  &	---  \\  
BVF262	    &	0.97  &	44.02 &	$-$23.28 & $-$23.33	& 44.73	& 0.29 & --- 	    &	---   &	---  &	---  \\  
PKS 0440$-$00 &	0.844 &	43.65 &	$-$23.92 & $-$23.81	& 44.93	& 0.41 & Mg~{\sc ii} &	50    &	0.18 &	0.33 \\
PKS 0938+18 &	0.943 &	43.94 &	$-$23.59 & $-$22.82	& 44.53	& 0.20 & Mg~{\sc ii} &	100   &	0.49 &	0.05 \\
3C~422	    &	0.942 &	43.94 &	$-$24.85 & $-$24.75	& 45.30	& 0.81 & Mg~{\sc ii} &	140   &	1.58 &	0.09 \\
MRC~2112+172  &	0.878 &	43.75 &	$-$24.85 & $-$23.58	& 44.83	& 0.35 & --- 	    &	---   &	---  &	---  \\  
4C~02.54    &	0.976 &	44.04 &	$-$23.91 & $-$25.11	& 45.45	& 1.0 & --- 	    &	---   &	---  &	---  \\  
\tableline
SGP2:11	    &	1.976 &	45.93 &	$-$23.30 & $-$24.40	& 45.16	& 0.63 & C~{\sc iv}  &	90    &	1.06 &	0.10 \\
SGP2:25	    &	1.868 &	45.78 &	$-$23.85 & $-$23.64	& 44.86	& 0.36 & C~{\sc iv}  &	70    &	0.61 &	0.08 \\
SGP2:36	    &	1.756 &	45.61 &	$-$23.78 & $-$23.12	& 44.65	& 0.25 & C~{\sc iv}  &	100   &	1.58 &	0.02 \\
SGP3:39	    &	1.964 &	45.91 &	$-$24.17 & $-$23.81	& 44.93	& 0.41 & C~{\sc iv}  &	85    &	1.06 &	0.06 \\
SGP4:39	    &	1.716 &	45.55 &	$-$21.85 & $-$24.18	& 45.07	& 0.53 & C~{\sc iv}  &	45    &	0.25 &	0.34 \\
PKS 1524$-$13 &	1.687 &	45.50 &	$-$24.15 & $-$24.96	& 45.39	& 0.94 & --- 	    &	---   &	---  &	---  \\
B2$-$2156+29  &	1.753 &	45.61 &	$-$25.75 & $-$25.17	& 45.47	& 1.10 & --- 	    &	---   &	---  &	---  \\
PKS 2204$-$20 &	1.923 &	45.85 &	$-$23.23 & $-$24.75	& 45.30	& 0.81 & --- 	    &	---   &	---  &	---  \\
4C~45.51    &	1.992 &	45.95 &	$-$26.24 & $-$25.97	& 45.79	& 1.97 & C~{\sc iv}/Mg~{\sc ii}  &  58/101 & 0.65/0.58 & 0.67  \\
\tableline
MZZ 9744    &	2.735 &	46.78 &	$-$23.84 & $-$24.15	& 45.05	& 0.52 & C~{\sc iv}  &	90    &	0.71 &	0.12 \\
MZZ 9592    &	2.71  &	46.76 &	$-$24.81 & $-$24.58	& 45.23	& 0.72 & C~{\sc iv}  &	90    &	0.63 &	0.19 \\
MZZ 1558    &	1.829 &	45.72 &	$-$23.01 & $-$24.11	& 45.05	& 0.51 & C~{\sc iv}  &	120   &	1.34 &	0.06 \\
MZZ 1140    &	1.735 &	45.58 &	$-$22.69 & $-$21.99	& 44.20	& 0.11 & C~{\sc iv}  &	50    &	0.21 &	0.05 \\
MZZ 4935    &	1.876 &	45.79 &	$-$21.75 & $-$22.02	& 44.21	& 0.11 & C~{\sc iv}  &	50    &	0.22 &	0.05 \\
\tableline
CTQ 414	    &	1.29  &	44.78 &	$-$22.95 & $-$24.40	& 45.13	& 0.60 & C~{\sc iv}/Mg~{\sc ii}  & 40/40   & 0.34/0.16 & 0.40  \\
\tableline
\enddata 
\tablecomments {\tiny 
Col. (1): Object name.  
Col. (2): Redshift.  
Col. (3): Distance modulus ($H_0=70$ km s$^{-1}$ Mpc$^{-1}$, 
          $\Omega_\Lambda=0.7$, $\Omega_{\rm m}=0.3$).  
Col. (4): Absolute $R$-band magnitude of the host.  
Col. (5): Absolute $V$-band magnitude of the quasar.  
Col. (6): Monochromatic $V$-band luminosity of the quasar (log base 10).  
Col. (7): \mbh\ derived using the \mbh-luminosity of quasars from
          Peterson et al. (2004) in units of $10^9 M_\odot$.  
Col. (8): Emission line used.
Col. (9): FWHM of line in Col. (8), measured manually in a ``double-blind''
manner.  The line width for PKS 0938+18 was measured by Jauncey et al. (1984);
PKS 0440$-$00 was measured by Jackson \& Browne (1991); 4C~45.51 was measured
by Stickel \& K\"uhr (1993); and 3C~422 was measured by Aldcroft, Bechtold, \&
Elvis (1994).
Col. (10): \mbh\ derived using C~{\sc iv} or Mg~{\sc ii} broad emission line
	   width and continuum luminosity.
Col. (11): Radiation efficiency in units of Eddington luminosity.
}
\end{deluxetable}

\clearpage

\begin{figure}
    \includegraphics[width=5.truein,height=7.truein,angle=-90]{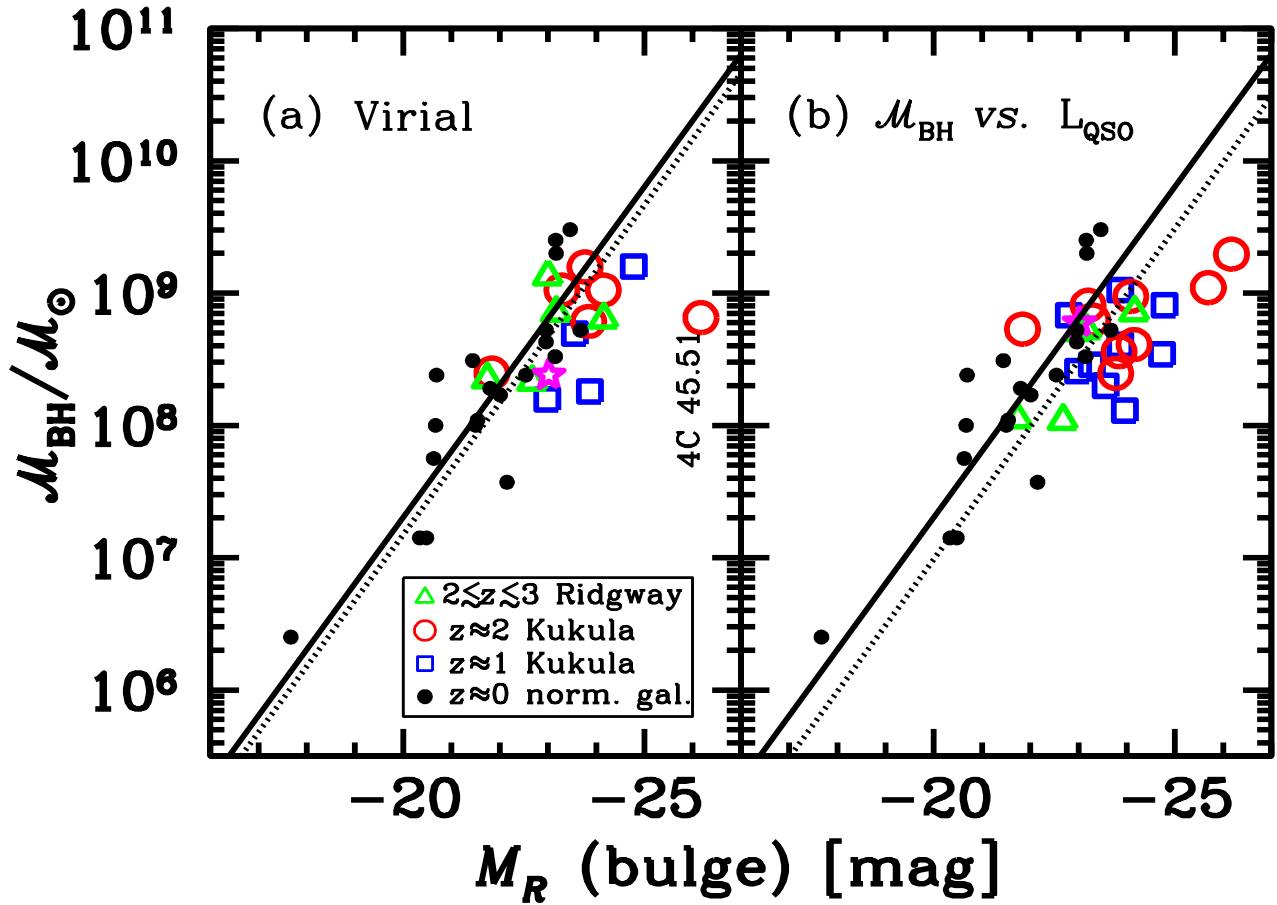}

    \caption {The relationship of the black hole mass, \mbh, vs. bulge
	      absolute luminosity, \mr, at low $z$ (solid round points;
	      compiled in Bettoni et al. 2003) and $z\gtrsim 1$ (open points).
	      Solid lines: line fitted to $z\approx0$ solid points.  Open
	      triangles: Ridgway et al.  (2001).  Open squares and Open
	      circles: Kukula et al.  (2001).  Open star: Peng et al.  (2005).
	      Panel ({\it a}):  \mbh\ derived using the virial mass estimate.
	      Panel ({\it b}):  \mbh\ derived assuming \mbh-luminosity
	      relation of quasars (Peterson et al. 2004).  The dotted lines
	      fitted to the $z\approx2$, radio-quiet AGN, points are
	      displaced from the $z=0$ relationship by 0.3 ({\it a}) and 0.6
	      ({\it b}) mag.}
\end{figure}


\newpage

\begin{figure}
    \plotone {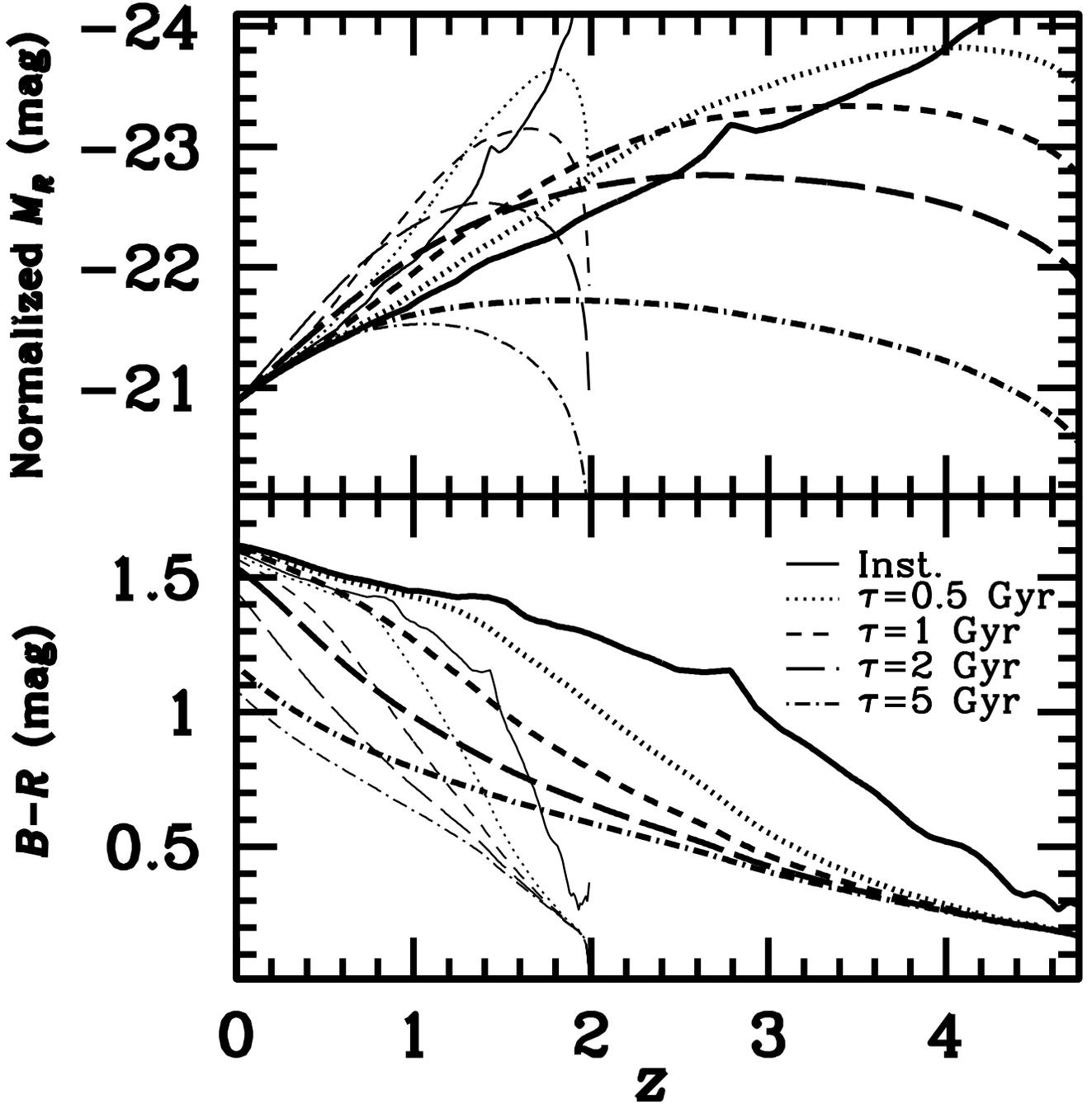}

    \caption {Luminosity evolution for simple star formation histories.  {\it
	      Top:} We plot the $R$-band luminosity evolution of stellar
	      populations as a function of redshift with formation redshifts
	      at $z_f = 2$ ({\it thin lines}) and $z_f = 5$ ({\it thick
	      lines}) for an instantaneous burst and $\tau$-model bursts
	      (Bruzual \& Charlot 2003). {\it Bottom:} We plot the colors of
	      the objects as a function of the corresponding star formation
	      histories.  The models are normalized to $M_R^* = -20.88$ (Brown
	      et al.  2001).}

\end{figure}


\begin{figure}

    \includegraphics[width=5.truein,height=7.truein,angle=-90]{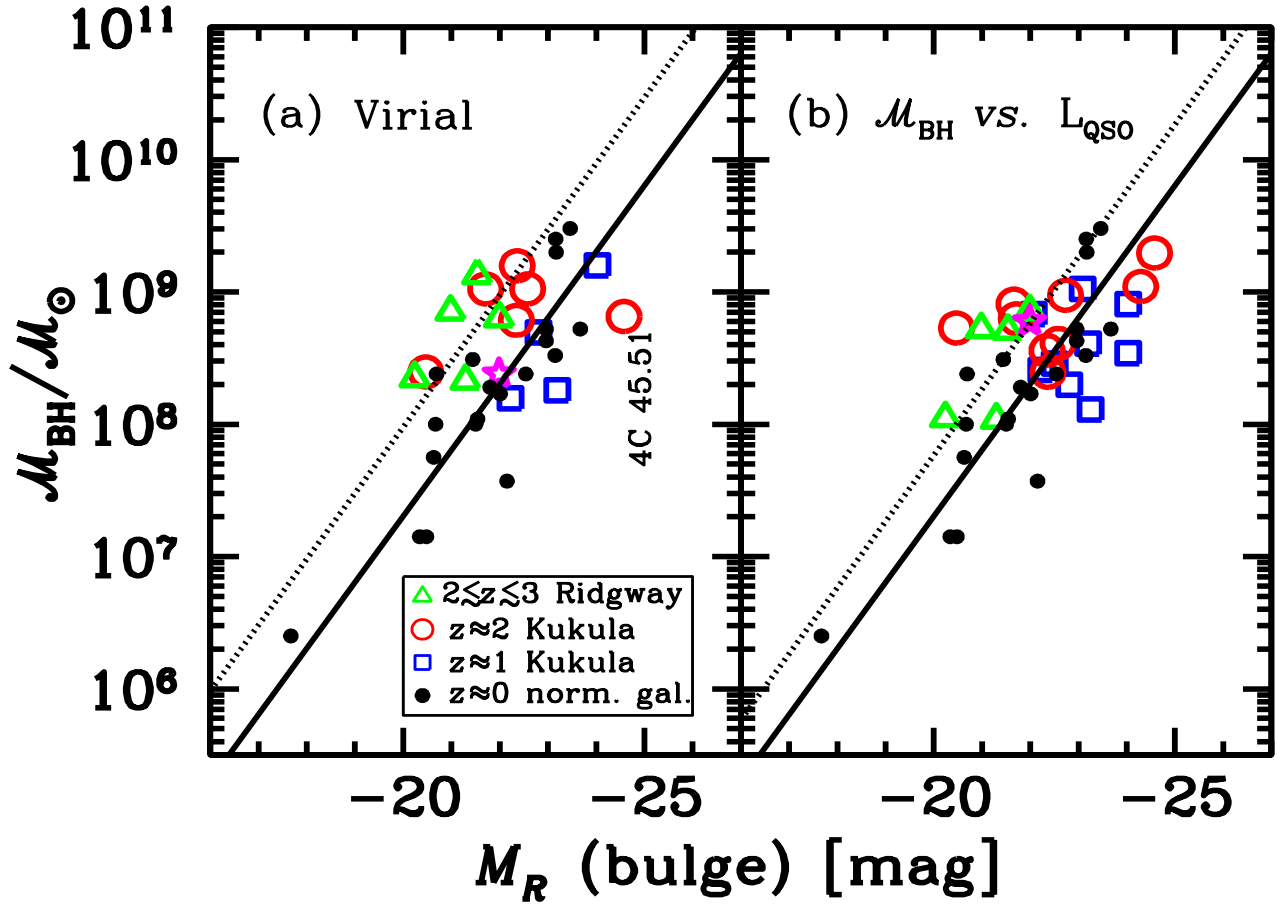}
    \caption {This plot is the same as Fig. 1, but the open points
	      are shifted horizontally by assuming that the hosts evolve {\it
	      passively} since $z=5$ by $d$\mr/$dz$ = $-0.8$ mag.  The dotted
	      lines fitted to the $z\approx2$, radio-quiet AGN, points are
	      displaced from the $z=0$ relationship by 1.3 ({\it a}) and 0.8
	      ({\it b}) mag.}
\end{figure}

\end {document}